\documentclass[aps,prb,twocolumn,showpacs,superscriptaddress]{revtex4-1}
\usepackage[mathscr]{eucal}
\usepackage{graphicx}
\usepackage{amsmath}
\usepackage{amstext}
\usepackage{amssymb}
\usepackage{amsfonts}
\usepackage{amsbsy} 
\usepackage{dcolumn}
\usepackage{bm}
\usepackage{multirow}
\usepackage{color}

\begin{document}
\title{Constructing symmetric topological phases of bosons in three dimensions \\
via fermionic projective construction and dyon condensation}
\author{Peng Ye}
\affiliation{Perimeter Institute for Theoretical Physics, Waterloo, Ontario, Canada N2L 2Y5}
\author{Xiao-Gang Wen}
\affiliation{Perimeter Institute for Theoretical Physics, Waterloo, Ontario, Canada N2L 2Y5}
\affiliation{
Department of Physics, Massachusetts Institute of Technology, Cambridge, Massachusetts 02139
}

\begin{abstract}
Recently, there is a considerable study on gapped symmetric phases of bosons
that do not break any symmetry.  Even without symmetry breaking, the bosons can
still be in many exotic new states of matter, such as symmetry-protected
trivial (SPT) phases which are short-range entangled  and symmetry-enriched
topological (SET) phases which are long-range entangled.  
It is well-known that \emph{non-interacting fermionic topological insulators}
are SPT states protected by time-reversal symmetry and U(1) fermion number
conservation symmetry.  In this paper, we construct three-dimensional exotic
phases of bosons with time-reversal symmetry and boson number conservation U(1)
symmetry by means of \emph{fermionic projective construction}. 
We first construct an \emph{algebraic bosonic insulator} which is a symmetric
bosonic state with an emergent U(1) gapless gauge field.  We then obtain 
many gapped
bosonic states that do not break the time-reversal symmetry and boson number
conservation via proper \emph{dyon condensations}.
We identify the constructed states by calculating the allowed electric and
magnetic charges of their excitations, as well as the statistics and the
symmetric transformation properties of those excitations.
This allows us to show that our constructed states can be trivial SPT states
(\emph{i.e.} trivial Mott insulators of bosons with symmetry), non-trivial SPT
states (\emph{i.e.} bosonic topological insulators) and SET states (\emph{i.e.}
fractional bosonic topological insulators). In  non-trivial SPT states, the
elementary monopole (carrying zero electric charge but unit magnetic charge)
and elementary dyon (carrying both unit electric charge and unit magnetic
charge) are fermionic and bosonic, respectively. In SET states, intrinsic
excitations may carry fractional charge.

\end{abstract}
 \maketitle
\tableofcontents

\section{Introduction}\label{Introduction}

A quantum ground state of a many-boson system can be in a
spontaneously-symmetry-breaking state, or a topologically ordered (TO)
state.\cite{Wtop,WNtop,Wrig} A TO state is defined by the following features:
ground state degeneracy in a topologically non-trivial closed
manifold,\cite{Wtop,WNtop,Wrig} or emergent fermionic/anyonic
excitations,\cite{H8483,ASW8422} or chiral gapless edge
excitations.\cite{H8285,Wedgerev}  If, in addition to a TO, the ground state
also has a symmetry, such a state will be referred as a ``symmetry-enriched
topological (SET) phase''.

Recently, it was predicted that even if the bosonic ground state does not break
any symmetry and has a trivial TO, it can still be in a non-trivial phase
called bosonic SPT phase.\cite{Chenlong,Chen10,Wenscience} Since the bosonic
SPT phases have only trivial TOs, a systematic description/construction of
those SPT phases were obtained via group cohomology
theory.\cite{Chenlong,Chen10,Wenscience} Many new SPT phases were
predicted/constructed with all possible symmetries and in any dimensions,
including three non-trivial bosonic SPT phases with  U(1) symmetry (particle
number conservation) and time-reversal symmetry (Z$^T_2$) in three dimensions.
We will refer those phases as bosonic topological insulators (BTI). If a SET
state with the same symmetry as BTI, we call it fractional BTI (\emph{f}BTI).
In the following, we also refer all gapped phases of bosons that do not break
the symmetry (including SPT and SET) as ``\emph{topological phases}''.

To realize bosonic TO phases or SPT phases, the interaction is crucial, since
without interaction, bosons always tend to condense trivially. This fact
hinders the perturbation approach if we want to realize TO or SPT phases.  One
useful approach is via the exactly soluble models, as in the string-net
approach\cite{FNS0428,LW0510}  and the group cohomology
approach.\cite{Chenlong,Chen10,Wenscience}  Recently, many other approaches
were proposed, which are based on field theory, topological invariants,
critical theory of surface, topological response theory
etc.\cite{LS0903,LV1219,VS1258,Burnell13,Cheng13,Chen13,Wen13topo,Xu13,Hung1211,Senthillevin12,Xu12,MM13}
A quite effective approach for strongly interacting systems is the ``projective
construction''.\cite{BZA8773,BA8880,AM8874,KL8842,SHF8868,AZH8845,DFM8826,WWZcsp,Wsrvb,LN9221,MF9400,WLsu2}
It has been recently realized that  the projective construction is also helpful
in constructing bosonic SPT states.\cite{GV1207,Lu12b,LL1263,YW12,MM13a}.
 
Roughly speaking, in the projective construction, the bosonic operator is split
into a product of parton operators. Different kinds of partons can individually
form different mean-field ground states. The physical ground state of the boson
system is realized by projecting the direct product of multiple mean-field
ground states into the physical Hilbert space $\mathscr{H}_{\rm phys.}$ in
which the multiple partons are glued together into a physical boson on each
site. In terms of path integral formulation, such a gluing process is done by
introducing fluctuating internal gauge fields that couple to partons.  

It is now well-known that three dimensional {\it non-interacting fermionic}
topological insulators (TI)\cite{TI} are classified by Z$_2$.  The free
fermionic TI state is protected by U(1)$\rtimes$Z$^T_2$. The trivial and
non-trivial TI phases can be labeled by the so-called ``axionic $\Theta$
angle'' in the electromagnetic response action ``$S_{\rm
EM}=\frac{\Theta}{8\pi^2}\epsilon^{\mu\nu\lambda\rho}\partial_\mu A_\nu
\partial_\lambda A_\rho$'' ($A_\mu$ is external electromagnetic potential).
$\Theta=0\ (\pi)$ corresponds to the trivial (non-trivial) phase.  It is
natural to ask whether there exists a bosonic version of TI, \emph{i.e.} BTI
and \emph{f}BTI via fermionic projective construction? Ref.
\onlinecite{Swingle10} applied the fermionic projective construction approach
in which the boson creation operator is split into a singlet pair of spin-1/2
fermions. It is assumed that the fermions are described by a non-trivial TI
mean-field ansatz which explicitly breaks the internal SU(2) gauge symmetry
down to $Z_2$. The resultant physical ground state is a SET state admitting a
fractional $\Theta$ angle and emergent Z$_2$ TO.  By definition, this bosonic
insulator is a $f$BTI, following Ref.  \onlinecite{Maciejko10} where a
fermionic version in the presence of strong interactions is proposed.
 
In the present work, we shall consider   U(1)$\rtimes$Z$^T_2$ symmetric
topological phases of bosons   in three dimensions.  We will use a fermionic
projective construction to construct symmetric gapless and gapped phases, the
latter of which may have non-trivial TO or SPT orders.  Our boson model
contains \emph{four} kinds of charge-1 bosons with U(1)$\rtimes$Z$^T_2$
symmetry in three dimensions (\emph{i.e.} Eq.(\ref{parton})). In this fermionic
projective construction, each boson is split into two different fermions
($f_1,f_2$) carrying ``spin-$1/2$''. $f_1$ and $f_2$ carry $\alpha$ and
$(1-\alpha)$ electric charge, respectively (see Table \ref{chargetable}).

To ensure that  before projection the mean-field ansatz of fermions respects
symmetry, we assume that  mean-field ansatzes of the fermions describe
fermionic gapped phase with $\theta$-angle $\theta_1$ for $f_1$ fermions and
$\theta_2$ for $f_2$ fermions.  We assume $\theta_1=\theta_2=0$ or
$\theta_1=\theta_2=\pi$ where the two fermions form the same trivial band
insulator state or the same topological insulator states.  We will use
$(\theta_1,\theta_2,\al)$ to label those  mean-field ansatzes.  Due to the
projective construction, an internal U(1) gauge field $a_\mu$ exists which is
gappless.  So our construction (at this first step) leads to gapless insulating
states of the bosons after the projection.  We call such a state algebraic
bosonic insulator.

To obtain gapped insulating states of the bosons, we shall push the internal
gauge field into its confined phases, where quantum fluctuations are very
strong which leads to a proliferation of certain dyon.  There are many
different kinds of confined phases that correspond to  proliferation of
different dyons.  These dyons may carry many quantum numbers including: fermion
numbers of $f_1,f_2$, magnetic charge and gauge charge of the internal gauge
field, magnetic charge and electric charge of external electromagnetic gauge
field.  We will use $(l,s)$ to label those different proliferated (or
condensed) dyons that do not break the U(1) and time reversal symmetries.

\begin{figure}[t]
\centering
\includegraphics[width=7cm]{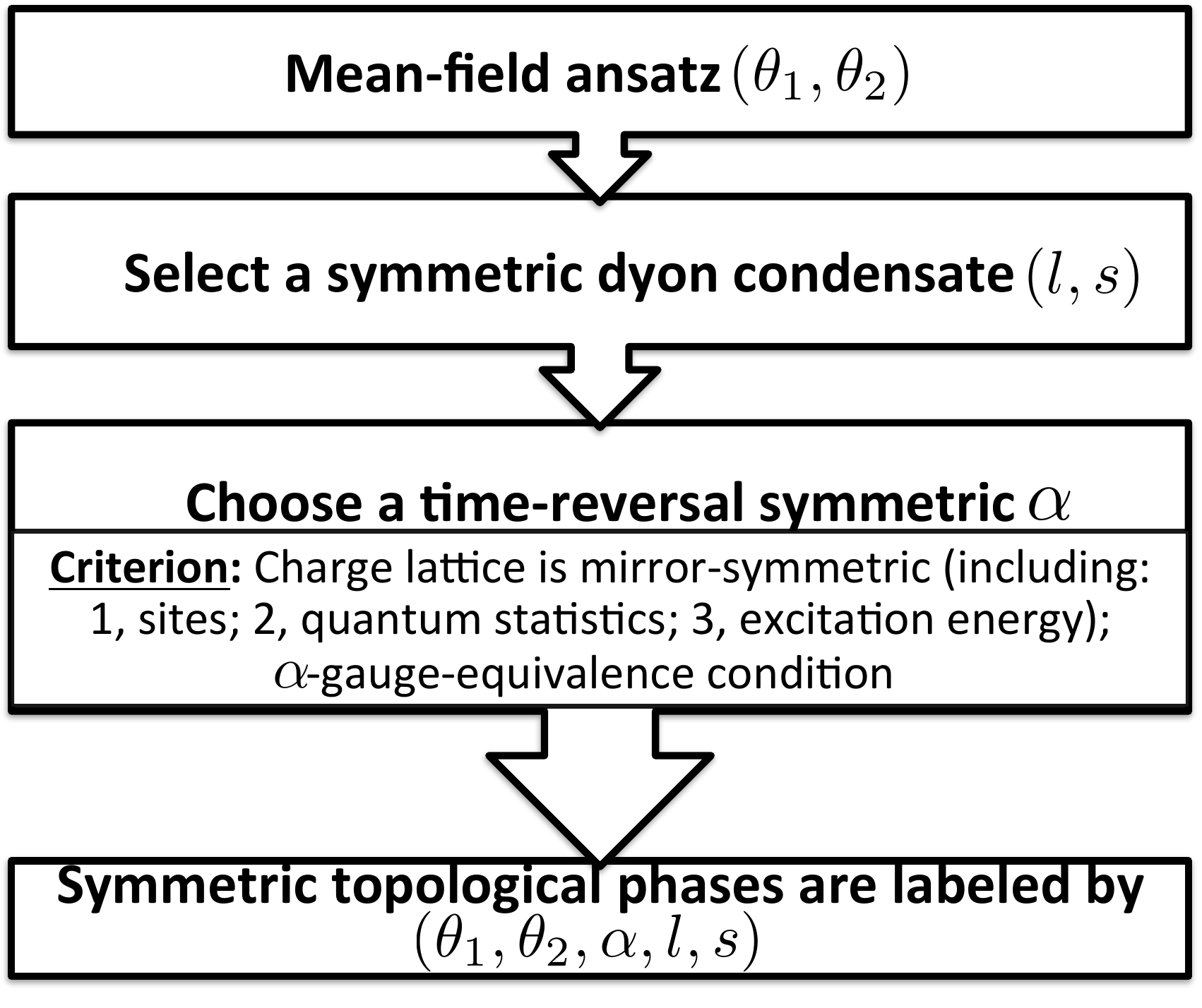}
\caption{Illustration of  the basic process of constructing symmetric topological phases via fermionic projective construction and dyon condensation.}
\label{figure_process}
\end{figure}

After selecting a symmetric dyon condensate $(l,s)$ and charge assignment
$\alpha$, we may construct a gapped topological phases with
U(1)$\rtimes$Z$^T_2$ symmetry.  We find that \emph{the dyon condensation break
the ``gauge symmetry'' of shifing $\al$ (dubbed ``$\alpha$-gauge symmetry''), so that different $\al$ will lead to
different bosonic states up to a gauge redundancy} (cf. Sec.\ref{axionfield}).  Thus, those   gapped topological phases are labeled
by $(\th_1,\th_2,\al)$ and $(l,s)$ (see Tables, \ref{tab:results0},
\ref{tab:results}, and \ref{tab:results1}), and we have to choose certain
special values of $\al$ ensure the U(1) and time reversal symmetries (see Fig.
\ref{figure_process}).  The excitation spectrum  above those dyon condensates
is formed by the so-called ``deconfined dyons'' that have trivial \emph{mutual
statistics} with the condensed dyon. 

The nature of topological phases in $\mathscr{H}_{\rm phys.}$ is determined by
the properties of \emph{intrinsic excitations} (defined as those excitations
that have zero magnetic charge of external electromagnetic field). A
topological phase where intrinsic excitations carrying fractional  electric
charge, fermionic statistics, or, other forms of fractionalization is a SET
state (\emph{i.e.} $f$BTI) with both TO and symmetry. If TO is absent and
symmetry is still unbroken, the topological phase must be a SPT state. If the
excitation spectrum of a SPT state admits non-trivial Witten effect with
$\Theta=2\pi$\cite{Witten79,RF10,MM13,VS1258,Ye13long}, the state is a non-trivial SPT
(\emph{i.e.} a BTI). Otherwise, the state is a trivial SPT, \emph{i.e.} a
trivial Mott insulator of bosons with symmetry. All topological phases that we
constructed are summarized in    Table \ref{tab:results} ($\alpha=1/2$) and
Table \ref{tab:results1} (for a general $\alpha$-sequence). These two tables
contain the general results. For reader's convenience, some concrete examples
of non-trivial BTI phases are shown in Table \ref{tab:results0}. The basic
process of constructing symmetric topological phases is shown in Fig.
\ref{figure_process}.

\begin{table}
\centering
\caption{Assignment of EM electric charge and $a_\mu$-gauge charge}
\label{chargetable}
\begin{tabular}{|c|c|c|c|}\hline
Particle&EM electric charge&$a_\mu$-gauge charge\\ \hline\hline
$f_1$&$\alpha$&$+1$\\ \hline
$f_2$&$1-\alpha$&$-1$\\ \hline
$b$&$+1$&$0$\\
\hline
\hline
\end{tabular}
\end{table}

 \begin{table}
 \begin{tabular}[t]{|c|c|c}
 \hline
 \begin{minipage}[t]{0.6in} $(\theta_1,\theta_2,\alpha)$\end{minipage} 
 & \begin{minipage}[t]{1.5in}  
 Dyon condensate: $(l,s)$\end{minipage}\\\hline \hline
 \begin{minipage}[t]{0.6in} $(0,0,1)$\end{minipage} 
 & \begin{minipage}[t]{1.5in} 
 $(1,1)$\end{minipage}\\\hline
 \begin{minipage}[t]{0.6in} $(0,0,1)$\end{minipage} 
 & \begin{minipage}[t]{1.5in} 
 $(3,1)$\end{minipage}\\\hline
 \begin{minipage}[t]{0.6in} $(\pi,\pi, -\frac{1}{2} )$\end{minipage} 
 & \begin{minipage}[t]{1.5in} 
 $(1,1)$\end{minipage}\\\hline
 \begin{minipage}[t]{0.6in} $(\pi,\pi, \frac{3}{2} )$\end{minipage} 
 & \begin{minipage}[t]{1.5in} 
 $(1,1)$\end{minipage}\\\hline
 \hline
  \end{tabular}
  \caption{Some concrete examples  of non-trivial BTI phases in three
dimensions. Each BTI state is labeled by five numbers
$(\theta_1,\theta_2,\alpha,l,s)$. $\theta_1$ ($\theta_2$) equals to $0$ or
$\pi$, denoting the trivial or non-trivial TI phases of the fermion $f_1$
($f_2$). $f_1$ and $f_2$ carry $\alpha$ and $(1-\alpha)$ electric charge of
external electromagnetic field, respectively.  In the
mean-field ansatz $(0,0)$, the condensed dyon is composed by $s$ magnetic
charge of internal U(1) gauge field and $l$ physical bosons; In the mean-field
ansatz $(\pi,\pi)$, the condensed dyon is composed by $s$ magnetic charge of
internal U(1) gauge field, $l$ physical bosons, and, in addition, $s$ $f_2$
fermions. Each physical boson is composed by one $f_1$ and one $f_2$. 
}
  \label{tab:results0}
\end{table}

The remaining parts of the paper are organized as follows. In
Sec.\ref{sec:boson}, the underlying boson degrees of freedom as well as the
fermionic projective construction is introduced. Symmetry operations (both
U(1) and Z$^T_2$) on physical bosons and fermonic partons are defined. In
Sec.\ref{sec:dyon}, the general properties of dyons are discussed. The main results of topological phases are derived in Sec.\ref{sec:82}
where topological phases are constructed by setting $\alpha=1/2$. The general  construction of topological phases in the presence of general $\alpha$-sequence is provided in Sec.\ref{sec:general}. Conclusions are made
in Sec.\ref{sec:conclusion}.

\section{Fermionic projective construction of many-boson state with
U(1)$\rtimes$Z$^T_2$ symmetry}\label{sec:boson}

\subsection{Definition of boson operators} 

We will use a system with four kinds of electric charge-1 bosons in three
dimensions.  Those bosons are described by four boson operators.  We split the
boson operators into two different spin-1/2 fermions:
\begin{align}
&(b_{1},b_{2}, b_{3},b_{4})=(f_{1\uparrow}f_{2\uparrow},f_{1\uparrow}f_{2\downarrow},f_{1\downarrow}f_{2\uparrow},f_{1\downarrow}f_{2\downarrow})\,.
\label{parton}
\end{align}
The fermionic projective construction of the four bosons implies that the underlying bosons are of hard-core nature since the dimension of the  bosonic Hilbert space at each lattice site-$i$ is truncated to be finite and exchange of two bosons at different sites do not generate fermionic sign. As a result, at the very begining, the underlying boson model on the lattice must be a correlated bosonic system. All possible ground states with boson charge conservation symmetry U(1) and time-reversal symmetry Z$^T_2$ are what we shall look for in this paper.  

The physical ground state wave function $|\text{GS}\rangle$ in $\mathscr{H}_{\rm phys.}$ can be written in terms of direct product of fermions' mean-field ansatz in subject to Gutzwiller projection:
\begin{align}
|\text{GS}\rangle=\hat{P}\Big(|\Psi(f_1)\rangle\otimes|\Psi(f_2)\rangle\Big)\,,
\end{align}
where, $\hat{P}$ is the Gutzwiller projection operator which enforces that the total number of $f_1$ is equal to that of  $f_2$ at each site in the physical (projected) Hilbert space $\mathscr{H}_{\rm phys.}$. $|\Psi(f_1)\rangle$ and $|\Psi(f_2)\rangle$ are mean-field ansatzes for the ground states of $f_1$ and $f_2$, respectively. In the present work, we assume that both of $f_1$ and $f_2$ form \emph{mean-field ansatzes} with band structures which respect U(1)$\rtimes$Z$^T_2$ symmetry. Such  band structures are of TI classified by Z$_2$, \emph{i.e.} one trivial state and one non-trivial state. It is thus instructive to separatly study the physical ground states in two different classes: 1, both are trivial; 2, both are non-trivial. Other mean-field ansatzes explicitly break time-reversal symmetry already at mean-field level.

 \subsection{Definition of symmetry transformations}\label{sec:sym_def}
 \subsubsection{Time-reversal symmetry}
 
  Under time-reversal, the above fermions transform as the usual spin-1/2
fermions, but with an additional exchange $f_{1\sigma}\leftrightarrow f_{2\sigma}$,
and the bosons transform as
\begin{align}
\label{btrns}
&b_{1}\rightarrow -b_{4}\,,b_{2}\rightarrow b_{2}\,,b_{3}\rightarrow b_{3}\,,
b_{4}\rightarrow -b_{1}\,.
\end{align}
For instance, $b_{2}=f_{1\uparrow}f_{2\downarrow}\stackrel{\textcircled{1}}{\longrightarrow}  -f_{1\downarrow}\,f_{2\uparrow}\stackrel{\textcircled{2}}{\longrightarrow}  -f_{2\downarrow}\,f_{1\uparrow}=f_{1\uparrow}f_{2\downarrow}=b_{2}$ such that $b_{2}$ is unchanged, where \textcircled{1} represents $f_{\downarrow}\rightarrow -f_{\uparrow},f_{\uparrow}\rightarrow f_{\downarrow}$ and \textcircled{2} represents exchange of labels: $1\leftrightarrow 2$. In this projective construction, there is an internal U(1) gauge
field, $a_\mu$ minimally coupled to $f_1,f_2$. The assignment of gauge charges carried by $f_s$ ($s=1,2$) is
shown in Table \ref{chargetable}. $f_1$ and $f_2$ carry +1 and $-1$ gauge charges of $a_\mu$, 
respectively, such that all physical boson operators are invariant under $a_\mu$ gauge transformation.  

\subsubsection{Boson number conservation U(1) and the charge assignement}
 
Each boson carries +1 fundamental electric charge of external electromagnetic
(EM) field $A_\mu$ such that one can make the following assignment for fermions
shown in Table \ref{chargetable}: $f_{1}$ and $f_2$ carry $\alpha$ and
$1-\alpha$ EM electric charge of $A_\mu$, respectively. Here, $\alpha$ is a
real number whose value should not alter the vacuum expectation value of EM
gauge-invariant operators.  More precisely, $\alpha$ is not a defining
parameter of the underlying boson model. Rather, it is introduced in the
projective construction at ultra-violet (UV) scale.  If we only change
$\alpha$, the projected wavefunction should not change, if the projection is
done \emph{exactly at lattice scale}.  So the physical properties should not
depend on $\al$.  

 \subsection{Residual $\alpha$-gauge symmetry after dyon condensation}\label{axionfield}
 

The above discussion about $\al$ suggests that $\alpha$ is a pure gauge degree
of freedom
(or more precisely: a gauge redundancy).
We conclude that:

\frm{
Before the dyon condensation,
there is a $\alpha$-gauge symmetry which is defined as:
$\alpha\rightarrow\alpha+\lambda$ where $\lambda$ is any real number. 
} 

Later we will see that the dyon condensation can break such an ``$\al$-gauge
symmetry'', just like the Higgs condensation can break the usual ``gauge
symmetry''.  However, we believe that dyon condensation does \emph{not} break
all the $\al$-gauge symmetry:

\frm{Shifting $\al$ by any integer remains to be a ``gauge symmetry'' even after
the dyon condensation.}

Equivalently, one may view that $\alpha$ forms a circle with unit perimeter.
The physical consideration behind this statement is that the \emph{EM charge
quantization} is unaffected by any integer shift at all, and, such an integer
shift is nothing but redefinition of field variables.  

In this paper, we will show that in the mean-field ansatz
$(\theta_1,\theta_2)=(0,0)$ case, all topological phases (including SPT and
SET) satisfy this statement (cf. Sec.\ref{sec:general}). In other words, after
$\alpha\rightarrow \alpha+1$, the calculated properties of the topological
phases are unaffected. 

However, the statement is invalid in the mean-field ansatz
$(\theta_1,\theta_2)=(\pi,\pi)$ via our continuum effective field theory approach (cf. Sec.\ref{sec:general}). After $\alpha\rightarrow\alpha+1$, The
physical properties are changed.
In this case, it appears that

\frm{$\alpha$-gauge symmetry with any \emph{odd} integer shift is broken by the
dyon condensation. Shifting $\al$ by any \emph{even} integer remains to be a
``gauge symmetry'', after the dyon condensation. }

%
%
%

At the moment, we do not understand why dyon condensations for
$(\theta_1,\theta_2)=(0,0)$ and $(\theta_1,\theta_2)=(\pi,\pi)$ lead to
differnt $\al$-gauge symmetry breaking. However, we would like to point out
that our field theoretic treatment on dyons is established in the continuum
limit of spacetime.  More rigorous approach, however, should involve the
regularization procedures of dyon fields on lattice. For example, dyons are not
point particles at all on lattice. Rather, dyons are regularized on a dual
hyper-cubic lattice of spacetime where magnetic charge and electric charge are
put on dual sublattices.  This more careful lattice consideration may allow us
to understand how dyons condensation may break the $\al$-gauge symmetry and in
which way the  $\al$-gauge symmetry is broken.  We leave this issue to future
work.


%

\section{General properties of dyons}\label{sec:dyon}

 \subsection{Quantum numbers of dyons}
 
 The projective construction is a very natural way to obtain topological phases with TO  since at the very beginning the fermionic degrees of freedom and internal gauge fields are introduced at UV scale.  To obtain
SPT phases, we must prohibit the emergence of TO, by {\it at
least} considering the confined phase of the internal gauge field, where, the dyons
of the internal gauge field play a very important role.  For the purpose of probing the EM response, the non-dynamical EM field is applied and is assumed to be compact. Thus, a  dyon
may carry gauge (electric) charges and magnetic charges of both
internal gauge field and EM field.  The terms ``gauge (electric) charge'' and ``magnetic charge'' are  belonging to both gauge fields, while, for the EM field, we specify the charges by adding ``EM''  to avoid confusion. A dyon can also include $f_1$ and $f_2$ fermions, resulting in nonzero ``fermion number''. Thus a generic dyon is
labeled by a set of quantum numbers that describe those gauge charges, 
magnetic charges and fermion numbers.  Specially, a {\it monopole} is defined as a special dyon which doesn't carry any kind of gauge (electric) charges.

 To describe those dyon excitations
systematically, let us assume that each fermion ($f_s$) couples to its own
gauge field ($A^{fs}_\mu$) with ``+1'' gauge charge.  In fact, $A^{fs}_\mu$ are
combinations of $a_\mu$ and $A_\mu$ (cf. Table \ref{chargetable}):
\begin{align}
A^{f1}_\mu\equiv a_\mu+ \alpha A_\mu\,\,,A^{f2}_\mu\equiv -a_\mu +(1-\alpha)A_\mu\,.
\end{align}
A dyon can carry the magnetic charges in  
$A^{fs}$ gauge groups, which are labeled by $\{N^{(s)}_{m}\}\in\mathbb{Z}$
$(s=1,2)$.  These two magnetic charges $\{N^{(s)}_{m}\}$ are related to the magnetic charge $N^{a}_m$ in $a_\mu$ gauge group and magnetic charge $N_M$ in $A_\mu$ gauge group in the following way:
\begin{align}
 N^{(1)}_m=N^{a}_m+ \alpha N_M\,, N^{(2)}_m=-N^{a}_m+(1-\alpha)N_M\,,\label{eq_quantization}
\end{align} 
where, the EM magnetic charge $N_M$ is integer-valued as usual: $N_M\in\mathbb{Z}$.  For this reason, the quantization of magnetic charge $N_m^a$ of $a_\mu$ gauge group is determined by  two integers ``$N^{(1)}_m$'' and ``$N_M$'' via Eq.(\ref{eq_quantization}) with a given $\alpha$.  The following relations are  useful:
\begin{align}
 N_M= N^{(1)}_m +N^{(2)}_m\,,\,\,N_m^a=(1-\alpha)N^{(1)}_m-\alpha N^{(2)}_m\,.\label{combine_NM}
\end{align}

A dyon can also carry the fermion numbers of $f_1,f_2$ denoted 
by $\{N_f^{(s)}\}, s=1,2$. They are related to magnetic charges in the
following way: 
\begin{align}
N^{(1)}_f=n^{(1)}_f+\frac{\theta_1}{2\pi}N^{(1)}_m\,,\,\,N^{(2)}_f=n^{(2)}_f+\frac{\theta_2}{2\pi}N^{(2)}_m\,,\label{eq:W}
\end{align}
where, the $\theta$-related
terms are {\it polarization electric charge clouds} due to Witten effect\cite{Witten79,RF10}
and $n^{(s)}_f$ are integer-valued, indicating that integer numbers of fermions
are able to be {\it trivially} attached to the dyon. The nature of ``polarization'' is related to the fact that this charge cloud doesn't contribute {\it quantum statistics} to dyons\cite{Goldhaber89}. $\theta_1$ and $\theta_2$ determine the topology of fermionic band structures of $f_1$ and $f_2$ respectively if symmetry group U(1)$\rtimes$Z$^T_2$ is implemented. For example, $\theta_1=0$ if $f_1$ forms a trivial TI ansatz and $\theta_1=\pi$ if $f_1$ forms a non-trivial TI ansatz.

 \subsection{Time-reversal transformation of dyons, gauge fields, and Lagrangians}\label{sec:time-reversal}

To see whether the ground state breaks symmetry or not, it is necessary to understand how the symmetry acts on dyon labels $(N^{a}_m\,,N_M\,,N_f^{(1)}\,,N_f^{(2)}, N^{(1)}_m,N^{(2)}_m)$ as well as gauge fields  (${A}_\mu\,,{a}_\mu\,, \widetilde{{A}}_\mu\,,\widetilde{{a}}_\mu$), where, $\widetilde{A}_\mu$ and $\widetilde{{a}}_\mu$ are two dual gauge fields which are introduced to describe the minimal coupling in the presence of magnetic charge.  

The fermion-exchange process defined in Sec.\ref{sec:sym_def} implies the following transformation rules obeyed by dyon labels (all transformed symbols are marked by ``$\overline{\,\,\,\,}$"):
\begin{align}
&\overline{N_f^{(1)}}=N_f^{(2)}\,,\overline{N_f^{(2)}}=N_f^{(1)}\,,\label{number1}
\\
& \overline{N_M}=-N_M\,,\overline{N_m^{a}}=N_m^{a}\,,
\label{number2}
\end{align}
where, Eq.(\ref{number1}) holds by definition. In Eq.(\ref{number2}), the EM magnetic charge's sign is reversed as usual, which is consistent to reverse the sign of the EM gauge potential $\mathbf{A}$ which is a polar vector:
\begin{align}
\overline{\mathbf{A}}=-\mathbf{A}\,.\label{gauge1}
\end{align}
The second formula in Eq.(\ref{number2}) can be understood in the following way.  A single $f_1$ fermion couples to $\mathbf{A}$ and $\mathbf{a}$ with $\alpha$ and $+1$ coupling constants respectively, as shown in Table \ref{chargetable}. A single $f_2$ fermion couples to $\mathbf{A}$ and $\mathbf{a}$ with $1-\alpha$ and $-1$ coupling constants respectively.  Under Z$^T_2$, all spatial components of gauge fields will firstly change signs and $\alpha$ is replaced by:
\begin{align}
\overline{\alpha}=1-\alpha\,.\label{alphabar}
\end{align}
 At this {\it intermediate} status, ${f_1}$ fermion couples to $\mathbf{A}$ and $\mathbf{a}$ with $-\overline{\alpha}$ and $-1$ coupling constants respectively, and, ${f_2}$ fermion couples to $\mathbf{A}$ and $\mathbf{a}$ with $-1+\overline{\alpha}$ and $+1$ coupling constants respectively.  The second step is to exchange the two fermions as defined in Sec.\ref{sec:sym_def}. By definition, $\overline{f_2}$ (\emph{i.e.} the new $f_2$ fermion after time-reversal transformation) should couple to $\overline{\mathbf{A}}$ and $\overline{\mathbf{a}}$ with $1-\alpha$ and $-1$ coupling constants respectively, which results in: $(1-\alpha)\overline{\mathbf{A}}-\overline{\mathbf{a}}=-\overline{\alpha}\mathbf{A}-\mathbf{a}$. Likewise, $\overline{f_1}$ couples to $\overline{\mathbf{A}}$ and $\overline{\mathbf{a}}$ with $\alpha$ and $+1$ coupling constants, such that, $\alpha\overline{\mathbf{A}}+\overline{\mathbf{a}}=(-1+\overline{\alpha})\mathbf{A}+\mathbf{a}$. Overall, we obtain the following rule by using Eq.(\ref{gauge1}):
\begin{align}
\overline{\mathbf{a}}=\mathbf{a}\,\label{gauge2}
\end{align}
which requires the relation of magnetic charges $\overline{N_m^{a}}=N_m^{a}$ in a self-consistent manner as shown in Eq.(\ref{number2}).  

Based on the above results, one may directly derive the transformation rules obeyed by other quantum numbers:
\begin{align}
&\overline{N_m^{(1)}}=-N_m^{(2)}\,,\overline{N_m^{(2)}}=-N_m^{(1)}\,,\nonumber\\
&\overline{n^{(1)}_f}=n^{(2)}_f+\frac{\theta_1+\theta_2}{2\pi}N^{(2)}_m\,,\nonumber\\
&\overline{n^{(2)}_f}=n^{(1)}_f+\frac{\theta_1+\theta_2}{2\pi}N^{(1)}_m\,,\label{change1}
\end{align}
Suppose that ${N_A}$ and $N^a$ are the bare EM electric charge and $a_\mu$-gauge charge carried by dyons (cf. Table \ref{chargetable}):
\begin{align}
N_A=\alpha N^{(1)}_f+(1-\alpha)N^{(2)}_f\,,N^a=N^{(1)}_f-N^{(2)}_f\,.\label{na}
\end{align}
We have: 
\begin{align}
\overline{N_A}=N_A\,,  \overline{N^{a}}=-N^{a}\,,\label{number3}
\end{align}
by noting that $\overline{\alpha}=1-\alpha$.

By definition, the curls of dual gauge potentials ($\widetilde{\mathbf{A}}\,,\widetilde{\mathbf{a}}$) contribute electric fields ($\mathbf{E},\mathbf{E}^{a}$). Therefore, the dual gauge potentials should obey the same rules as electric fields under time-reversal transformation. And electric fields should also obey the same rules as electric charges ($N_A, N^a$) in a consistent manner such that the dual gauge potentials are transformed in the following way:
\begin{align}
\overline{\widetilde{\mathbf{A}}}=\widetilde{\mathbf{A}}\,\,,\overline{\widetilde{\mathbf{a}}}=-{\widetilde{\mathbf{a}}}\,.\label{gauge4}
\end{align}
The four formulas in Eqs.(\ref{gauge1}), (\ref{gauge2}), (\ref{gauge4}) are transformation rules obeyed by the spatial components of the gauge potentials.  The time-components of the gauge potentials (${{A}}_0\,,{{a}}_0\,,\widetilde{{A}}_0\,,\widetilde{{a}}_0$) obey the following rules:
\begin{align}
&\overline{A_0}=A_0\,,\overline{a_0}=-a_0 \,,\\
&\overline{\widetilde{A}_0}=-\widetilde{A}_0\,,\overline{\widetilde{a}_0}=\widetilde{a}_0\,\label{gauge5}
\end{align}
by adding an overall minus sign in each of  Eqs.(\ref{gauge1}), (\ref{gauge2}), and (\ref{gauge4}).

On the other hand, let us  consider the effective Lagrangian which describes the dyon dynamics. Let us  start with a general dyon $\phi$ and try to understand its time-reversal partner $\overline{\phi}$. The effective Lagrangian term $\mathscr{L}^K$ which describes the kinetic energy of $\phi$ can be written as:
\begin{align}
\mathscr{L}_{\rm kin}[\phi]=&\frac{1}{2m}|(-i\nabla+N^{a} \mathbf{a}+N_{A} \mathbf{A} + N^{a}_m \widetilde{\mathbf{a}}+N_M \widetilde{\mathbf{A}})\phi|^2 .
\end{align}
Here, we are performing time-reversal transformation in field theory action such that we keep all real-valued numerical coefficients ($N^{a},N_A, \dots$) but change all field variables. By using Eqs.(\ref{gauge1},\ref{gauge2},\ref{gauge4},\ref{gauge5})  and noting that $\overline{-i\nabla}=i\nabla$, we obtain the result:
\begin{align}
\mathscr{L}_{\rm kin}[\overline{\phi}]=&\frac{1}{2m}|(i\nabla+{N^{a}}\overline{ \mathbf{a}}+{N_{A}} \overline{\mathbf{A}}+ {N^{a}_m}\overline{ \widetilde{\mathbf{a}}}+{N_M }\overline{\widetilde{\mathbf{A}}})\overline{\phi}|^2\nonumber\\
=&\frac{1}{2m}|(-i\nabla-N^{a} \mathbf{a}+N_{A} \mathbf{A}+N^{a}_m \widetilde{\mathbf{a}}-N_M \widetilde{\mathbf{A}})\overline{\phi}|^2 \,.
\label{Lagrangian2}
\end{align}
 
Time component is similar:
\begin{align}
\mathscr{L}_{\rm t}[\phi]=\frac{1}{2m}|(i\partial_t+N^{a} {a}_0+N_{A} {A}_0+ N^{a}_m \widetilde{{a}}_0+N_M \widetilde{{A}}_0)\phi|^2
.\end{align}
After Z$^T_2$ operation,
\begin{align}
\mathscr{L}_{\rm t}[\overline{\phi}]=&\frac{1}{2m}|(i\partial_t-N^{a} {a}_0+N_{A} {A}_0+ N^{a}_m \widetilde{{a}}_0-N_M \widetilde{{A}}_0)\overline{\phi}|^2.\label{Lagrangian2}
\end{align}

\subsection{Mutual statistics and quantum statistics}\label{sec:mutual}

One of important properties of dyons is their 3D ``mutual statistics''. Two dyons with different quantum numbers may perceive a nonzero
quantum Berry phase mutually. More specifically, Let us  fix one dyon (``$\phi_1$'') at
origin and move another dyon ``$\phi_2$'' (labeled by symbol with primes) along a
closed trajectory which forms a solid angle $\Omega$ with respect to the
origin. Under this circumstance, one can calculate the Berry phase  that is added into the single-particle wavefunction of $\phi_2$:
\begin{align}
\text{Berry phase}=\frac{1}{2}\left[\sum_{s}N^{(s)}_m{N^{(s)}_f}'-\sum_s{N^{(s)}_m}'N^{(s)}_f\right]\Omega\,.\nonumber
\end{align}
If the Berry phase is nonvanishing for \emph{any} given $\Omega$, \emph{i.e.}
$\sum_{s}N^{(s)}_m{N^{(s)}_f}'\neq\sum_s{N^{(s)}_m}'N^{(s)}_f$, these two dyons
then have a non-trivial ``mutual statistics''. The physical consequence of mutual statistics is the following. If the confined phase of the internal gauge field is formed by a condensate of dyon $\phi_1$, all other allowed deconfined particles (\emph{i.e.} the particles which may form the excitation spectrum with a finite gap) must have trivial mutual statistics with respect to $\phi_1$, \emph{i.e.}
\begin{align}\sum_{s}N^{(s)}_m{N^{(s)}_f}'=\sum_s{N^{(s)}_m}'N^{(s)}_f\,.\label{trivialm1}
\end{align}
Otherwise,  they are confined by infinite energy gap.  There are two useful corollaries: (\emph{i}) It is obvious that a particle has a trivial mutual statistics with respect to itself; (\emph{ii})  We also note that $N^{(s)}_f$ and ${N^{(s)}_f}'$  may be replaced by integers $n^{(s)}_f$ and ${n^{(s)}_f}'$, respectively, by taking Eq.(\ref{eq:W}) into consideration. As a result, the criterion of {\it trivial mutual statistics} Eq.(\ref{trivialm1}) may be equivalently expressed as
\begin{align}
 \sum_{s}N^{(s)}_m{n^{(s)}_f}'=\sum_s{N^{(s)}_m}'n^{(s)}_f\,. \label{trivialm2}
\end{align}

On the other hand, it is also crucial to determine the quantum statistics of a generic dyon.  A generic dyon can be
viewed as $N_m^{(s)}$  magnetic charges of $A^{fs}_\mu$ gauge field attached by
$n^{(s)}_f$ $f_s$ fermions.  The quantum statistics of such a dyon is given by
\begin{align}
\text{Sgn}=\prod_{s}(-1)^{N_m^{(s)}n^{(s)}_f}(-1)^{n^{(s)}_f}\,, \label{quantumsign}
\end{align}
where $+/-$ represents bosonic/fermionic.\cite{Goldhaber89}  The first part,
$(-1)^{N_m^{(s)}n^{(s)}_f}$, is due to the interaction between the magnetic
charge $N_m^{(s)}$ and the gauge charge $n^{(s)}_f$ of the dyon. The polarization electric charges due to Witten effect do not attend the formation of internal angular momentum of electric-magnetic composite according to the exact proof by Goldhaber \emph{et.al.} \cite{Goldhaber89} so that $n^{(s)}_f$ instead of $N_f^{(s)}$ is put in Eq.(\ref{quantumsign}). One may also express $n^{(s)}_f$ in terms of ``$N_f^{(s)}-\frac{\theta_s}{2\pi}N^{(s)}_m$''. After this replacement,  it should be kept in mind that both ``$N_f^{(s)}-\frac{\theta_s}{2\pi}N^{(s)}_m$'' and ``$N^{(s)}_m$'' are integer-valued and $N_f^{(s)}$ can be any real number in order to ensure that $n_f^{(s)}$ are integer-valued. The second
part, $(-1)^{n^{(s)}_f}$, is due to the Fermi statistics from the attachment of
$n^{(s)}_f$ $f_s$ fermions.  Alternatively, the quantum statistics formula (\ref{quantumsign}) can be  reorganized into the following $\text{Sgn}\equiv (-1)^{\Gamma}$.
\begin{align}
{\Gamma}\doteq~\Gamma_1+\Gamma_2+\Gamma_3\label{eqn:stat1}
\end{align}
with 
\begin{align}
\Gamma_1\doteq N_M(\alpha n^{(1)}_f+(1-\alpha)n^{(2)}_f)\,&\,, \, \Gamma_2\doteq N^{a}_m(n^{(1)}_f-n^{(2)}_f)\,,\nonumber\\
\Gamma_3\doteq n^{(1)}_f+n^{(2)}_f&\,\nonumber
\end{align}
in which $\Gamma_1,\Gamma_2$ are contributed from the two gauge groups $A_\mu$ and $a_\mu$, respectively. $\Gamma_3$ is from fermionic sign carried by the attached fermions. The notation ``$\doteq$'' here represents that the two sides of the equality can be different up to any even integer.

\section{Topological phases with symmetry: $\alpha=1/2$} \label{sec:82}
  \subsection{Algebraic  bosonic insulators: Parent states of topological phases}\label{sec:abi}
 
 Let us
use the projective construction to study an exotic gapless bosonic
insulator (without the dyon condensation) which is called ``Algebraic bosonic
insulator (ABI)'' and can be viewed as a {\it parent state} of gapped symmetric
phases (\emph{i.e.} SPT and SET phases).  

For keeping time-reversal symmetry at least at mean-field level, we will only focus on $(\theta_1,\theta_2)=(0,0)$ and $(1,1)$. The ABI state does not break the U(1)$\rtimes$Z$^T_2$ symmetry since all
possible dyons (each dyon and its time-reversal partner) are included without
condensation. But the bulk is gapless since it contains an emergent gapless
U(1) gauge boson described by $a_\mu$.    The emergent U(1) gauge bosons are
neutral.  In addition to the emergent U(1) gauge bosons, ABI also contains many
dyon excitations, which may carry fractional electric charges ($N_f^{(s)}$ in
Eq.(\ref{eq:W})) and emergent Fermi statistics (determined by
Eq.(\ref{quantumsign})).  Since all electrically charged excitations are
gapped, such a phase is an electric insulator. 
At mean-field level, if $\theta_1=\theta_2=\pi$,  a
gapless surface state emerge which is described by Dirac fermions. Beyond
mean-field theory, those  gapless surface Dirac fermions (possibly with Fermi
energies aways from the nodes) will interact with the emergent U(1) gauge
fields which live in 3+1 dimensions. 

We note that the  internal U(1) gauge field $a_\mu$ has strong quantum fluctuations, and its ``fine structure
constants'' are of order 1.  It is  possible (relying on the physical boson Hamiltonian) that
the  internal U(1) gauge field is driven into a confined phase of gauge theory due to too strong
quantum fluctuations.  Due to the strong quantum fluctuations, the internal
U(1) gauge field configuration will contain many monopoles and even more
general dyons. The ABI discussed above is realized as an
unstable gapless fixed-point residing at the boundary between Coulomb phase and confined phase. ABI finally flows into a strongly-coupled fixed-point of confined phase by energetically condensing a bosonic dyon and thus opening  a bulk gap. In this case, it is possible that some non-trivial topological (gapped) phases with a global symmetry (including SPT and SET states) may be constructed in this confined phase that is featured by dyon condensations. As we have seen that our ABI has many kinds of dyons.  Relying on the details
of the physical boson Hamiltonian, different dyon condensations may appear. 
Different dyon condensations will lead to many different confined phases.

In the following, we  will set $\alpha=1/2$ and focus on looking for dyon condensations that generate a bulk spectral gap, and, {\it most importantly}, respect U(1)$\rtimes$Z$^T_2$ symmetry. All topological phases  are summarized in Table \ref{tab:results}. We should note that in each mean-field ansatz, only one dyon whose quantum numbers are self-time-reversal invariant is condensed to form a topological phase. As a matter of fact, two time-reversal conjugated dyons can condense simultaneously, still without breaking time-reversal symmetry. But this situation is  trivially back to the single dyon condensate for the reason that the two dyons are exactly same once the trivial mutual statistics between them is considered. (The details can be found in Appendix \ref{appendix_twodyons}).

\subsection{Standard Labeling and defining properties of topological phases}\label{sec:definingwitten}

Before moving on to topological phases of boson systems, we need to quantitatively define trivial SPT, non-trivial SPT and SET states based on physically detectable properties in EM thought experiments (compactness of EM field is assumed).

Each dyon is sufficiently determined by four independent quantum numbers in ABI state. The total number of independent quantum numbers will be decreased to three in a speficic topological phase where the condensed dyon  provides a constraint on the four quantum numbers as we will see later. There are many equivalent choices of labelings. In the following, we choose $(N_M,N^{(1)}_m,n^{(1)}_f,n^{(2)}_f)$ these four integer-valued quantum numbers to express the final key results of a given mean-field ansatz, such as quantum statistics and the total EM electric charge of excitations. We call it ``\emph{Standard Labeling}''. Basing on these four integers, we can obtain $N^{(2)}_{m}$, $N^{(1)}_f$ and $N^{(2)}_f$ via Eqs.(\ref{combine_NM},\ref{eq:W}).
As a result, $N_A$ and $N^a$ can be determined by Eq.(\ref{na}). In each mean-field ansatz, we will unify all key results  by using the \emph{Standard Labeling}.

 A trivial SPT state has the following properties:
\begin{enumerate}
\item  
Quantum statistics: $\Gamma_1=N_M N_E$.
\item Quantization condition: (i) $N_M\in\mathbb{Z}$\,, $N_E\in\mathbb{Z}$; (ii) At least one excitation exists for any given integer combination $(N_M, N_E)$.
\item TO doesn't exist.
\end{enumerate} 
The first two conditions (quantum statistics plus quantization condition) define a ``\emph{charge lattice}'' formed by two discrete data $N_M$ (y-axis) and $N_E$ (x-axis).  Differing from $N_A$ which is ``bare EM electric charge'', $N_E$ is the ``{\it total EM electric charge}'' in which possible dynamical  screening arising from the ground state is taken into consideration. In an experiment, $N_E$ is detectable while  $N_A$ is not.

\begin{figure}[t]
\centering
\includegraphics[width=8.5cm]{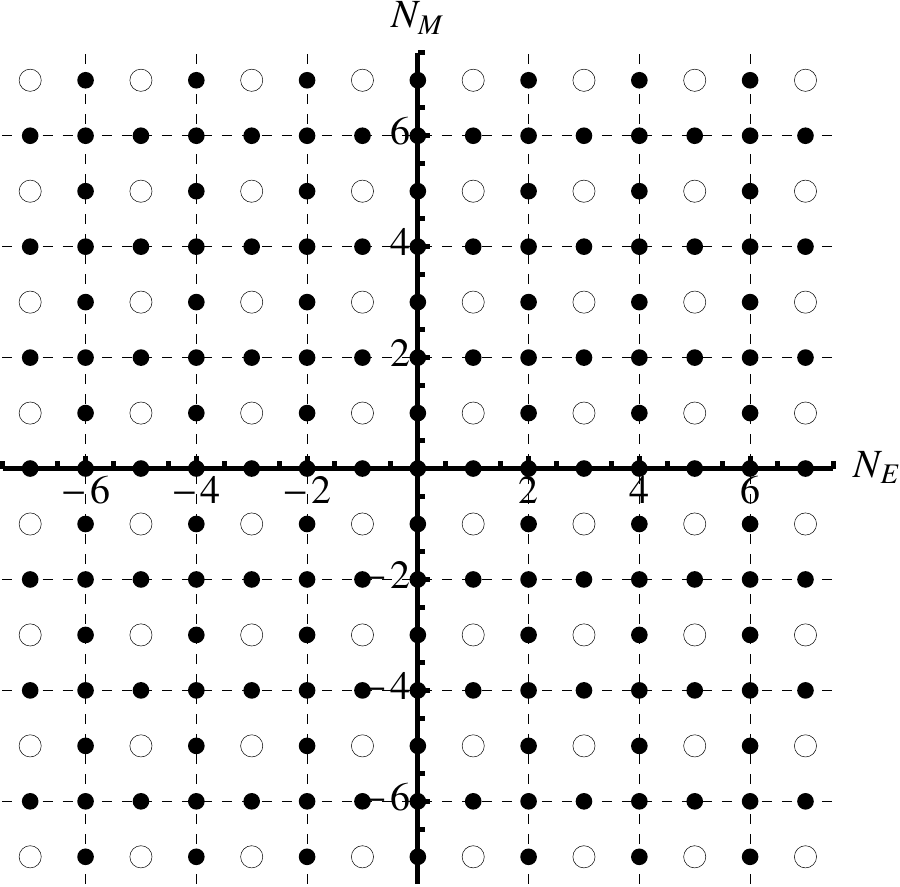}
\caption{Illustration of the charge lattice of trivial SPT. Excitations with $-7\leq N_E\leq 7$ and $-7\leq N_M\leq 7$ are plotted. The solid circles (open circles) denote bosonic (fermionic) statistics.  The  {elementary EM monopole} ($N_M=1,N_E=0$) is bosonic while the  {elementary EM dyon} ($N_M=1,N_E=1$) is fermionic.}
\label{figure_chargelattice0}
\end{figure}

In the trivial SPT state here, the charge lattice is corresponding to the phenomenon ``trivial Witten effect''. It rules out TO with fractional electric charges for intrinsic excitations and TO with fermionic intrinsic excitations. The trivial Witten effect implies the \emph{elementary EM monopole} ($N_M=1,N_E=0$) is bosonic while the \emph{elementary EM dyon} ($N_M=1,N_E=1$) is fermionic). This charge lattice is shown in Fig.\ref{figure_chargelattice0}. 

\begin{figure}[t]
\centering
\includegraphics[width=8.5cm]{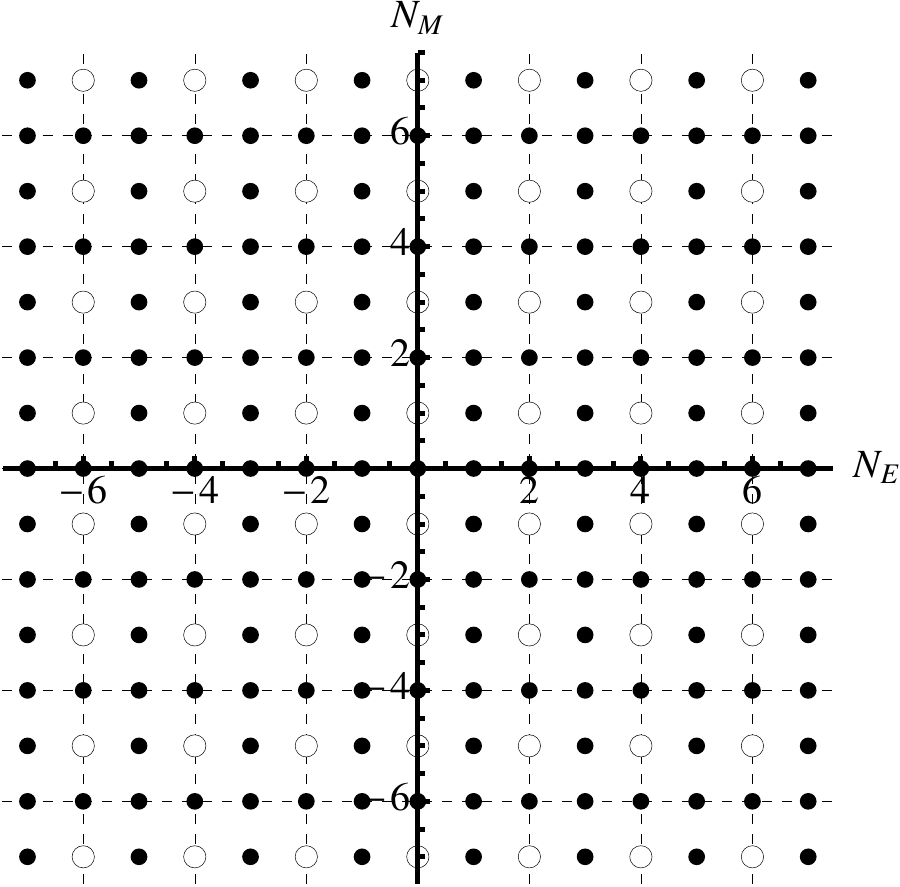}
\caption{Illustration of the charge lattice of non-trivial SPT. Excitations with $-7\leq N_E\leq 7$ and $-7\leq N_M\leq 7$ are plotted. The solid circles (open circles) denote bosonic (fermionic) statistics.  The  {elementary EM monopole} ($N_M=1,N_E=0$) is fermionic while the  {elementary EM dyon} ($N_M=1,N_E=1$) is bosonic.}
\label{figure_chargelattice2pi}
\end{figure}

A non-trivial SPT state  has the following properties:
\begin{enumerate}
\item  
Quantum statistics: $\Gamma_{2}=N_M (N_E-N_M)\doteq N_M N_E-N_M$.
\item Quantization condition: (i) $N_M\in\mathbb{Z}$\,, $N_E\in\mathbb{Z}$; (ii) At least one excitation exists for any given integer combination $(N_M, N_E)$.
\item TO doesn't exist.
\end{enumerate}
The first two conditions here correspond to the charge lattice with the phenomenon ``non-trivial Witten effect with $\Theta=2\pi\,\,\text{mod}(4\pi)$''. It also rules out TO with fractional electric charges for intrinsic excitations and TO with fermionic intrinsic excitations. The non-trivial Witten effect implies the \emph{elementary EM monopole} ($N_M=1,N_E=0$) is fermionic while the \emph{elementary EM dyon} ($N_M=1,N_E=1$) is bosonic). This charge lattice is shown in Fig.\ref{figure_chargelattice2pi}. This statistical transmutation has been recently discussed in Ref. \onlinecite{MM13}.

If the state \emph{with symmetry} supports a charge lattice which cannot be categorized into both of trivial and non-trivial SPT state, it must be a SET state.

 A constructed topological phase is a time-reversal symmetric state if the following three conditions are satisfied:
\begin{itemize}
\item \textbf{\fbox{Condition-\uppercase\expandafter{\romannumeral1}
}: The dyon condensate is Z$^T_2$-symmetric}. The selected condensed dyon is self-time-reversal symmetric (time-reversal pair condensates are not possible, see Appendix \ref{appendix_twodyons}.)
\item   \textbf{\fbox{Condition-\uppercase\expandafter{\romannumeral2}
}: The charge lattice is mirror-symmetric about x-axis}. On the charge lattice, the distribution of sites, quantum statistics and excitation energy   are mirror-symmetric about x-axis. More specifically, ($N_E,N_M$) and ($N_E,-N_M$) are simultaneously two sites of the charge lattice. At each site there are many excitations which are further labeled by the third quantum number (\emph{e.g.} $N^{(1)}_m$) in addition to the given $N_E$ and $N_M$. Each excitation ($N_E,N_M,N_m^{(1)}$) has a counterpart ($N_E,-N_M,N_m^{(1)'}$) with the same quantum statistics and the same excitation energy, and \emph{vice versa}. 

\item \textbf{\fbox{Condition-\uppercase\expandafter{\romannumeral3}
}: $\alpha$-gauge equivalence condition}. $2\alpha=\text{integer}$ in the mean-field ansatz $(0,0)$;  $\alpha=\text{half-odd}$ in the mean-field ansatz $(\pi,\pi)$. This condition and Condition-\uppercase\expandafter{\romannumeral2}  determine $\alpha$ altogether. Details of Proof  and related discussions on this condition are present in Sec.\ref{axionfield} and Sec.\ref{sec:general}.

\end{itemize} 
We simply say that ``\emph{the charge lattice is mirror-symmetric}'' if Condition-\uppercase\expandafter{\romannumeral2}  is satisfied. These three conditions lead to time-reversal invariance of the whole excitation spectrum.

 \subsection{Mean-field ansatz $(\theta_1,\theta_2)=(0,0)$}\label{sec:0+0}

 \subsubsection{Dyon condensation with symmetry}
Let us  first consider the most simplest starting point: the mean-field ansatz with $(\theta_1,\theta_2)=(0,0)$. In other words, both of fermions ($f_1,f_2$) are trivial TI. In this case, 
\begin{align}
(N^{(1)}_f)_c=(n^{(1)}_f)_c\in\mathbb{Z}\,,(N^{(2)}_f)_c=(n^{(2)}_f)_c\in\mathbb{Z}
\end{align}
according to Eq.(\ref{eq:W}). Hereafter, we use the subscript ``$c$'' to specify all symbols related to the condensed dyon ``$\phi_c$''. Excitations $\phi$ are labeled by symbols without subscript $c$. Thus, in the present mean-field ansatz $(\theta_1,\theta_2)=(0,0)$, the quantum numbers of excitations take values in the followng domains:  
 \begin{align}
&N^{(1)}_f=n^{(1)}_f\in\mathbb{Z}\,,N^{(2)}_f=n^{(2)}_f\in\mathbb{Z}\,,\label{alwaysinteger}\\
&N^a\equiv N^{(1)}_f-N^{(2)}_f=n^{(1)}_f-n^{(2)}_f\in\mathbb{Z}\,.\label{alwaysinteger1}
\end{align}
 Condition-\uppercase\expandafter{\romannumeral1}  further restricts $(N^{(1)}_f)_c=(N^{(2)}_f)_c$. Therefore, a general dyon with time-reversal symmetry is labeled by two integers $l$ and $s$, \emph{i.e.} 
\begin{align}
(N^a_m)_c=s\,,(N^{(1)}_f)_c=l\,,(N^{(2)}_f)_c=l\,,(N_M)_c=0\,.
\end{align}
 According to Eq.(\ref{eq_quantization}) where $(N_m^{(1)})_c\in\mathbb{Z}$ and $(N_M)_c=0$, $(N^a_m)_c$ is also an integer. Such a time-reversal symmetric dyon is always bosonic since $(\Gamma)_c=$even integer according to Eq.(\ref{eqn:stat1}). Most importantly, by definition, the EM field here is a probe field such that once the EM field is switched off, the physical ground state (formed by the dyon considered here) should not carry EM magnetic charge.  Therefore, $(N_M)_c$ vanishes, which is also required by time-reversal symmetry. Other quantum numbers of the condensed dyon are straightforward: 
\begin{align}
&(N_A)_c=l\,,(N^{(1)}_m)_c=s\,,(N^{(2)}_m)_c=-s\,,\nonumber\\
&(N^a)_c=0\,,(n^{(1)}_f)_c=l\,,(n^{(2)}_f)_c=l.
\end{align}


 But will such a dyon condensed state respect the EM electric U(1) symmetry and behave like
an insulator? To answer this question, according to Sec.\ref{sec:time-reversal}, let us write down the effective Lagrangian of the
condensed dyon $\phi_c$ in real time (only spatial components are written here for simplicity and time component is similar):
\begin{align} 
\label{Lagrangian}
 \mathscr{L}_{\rm kin}[\phi_c]&=\frac{1}{2 m}| (-i\nabla+s\widetilde{\mathbf{a}}+l\mathbf{A})\phi_c |^2
-V(\phi_c)\,,
 \end{align}
where, $V(\phi_c)$ is a symmetric potential energy term which energetically stabilizes the bosonic condensate.

In the dyon condensed state $\phi_c\neq 0$,
the internal gauge field $a_\mu$ is gapped and satisfies 
\begin{align}
s\widetilde{\mathbf{a}}=-l\mathbf{A}\,\label{aAaA}
\end{align} 
which indicates that the internal gauge field cannot fluctuate freely and is locked to the non-dynamical EM background.  We see that the
dyon condensation does not generate the $\mathbf{A}^2$ term if
\begin{align}
s\neq 0\,.\label{s0}
\end{align}
This requirement may be understood in the following way. If $s=l=0$, there is no dyon condensation, which is nothing but the ABI state discussed in Sec.\ref{sec:abi}. If $s=0$ and $l\neq0$, Eq.(\ref{aAaA}) reduces to $\mathbf{A}=0$ which is the consequence of mass term $\mathbf{A}^2$, a fingerprint of superconductor/superfluid with broken U(1). This case is nothing but condensation of $l$-bosons (carrying EM electric charge $l$), which breaks U(1) symmetry down to Z$_{|l|}$ discrete symmetry spontaneously ($Z_1$ represents complete breaking of U(1)).   
In the following, in order to preserve U(1) symmetry and consider dyon condensation, we will restrict our attention to $s\neq 0$. Thus the  dyon
condensed state indeed respects the EM electric U(1) symmetry and represents a fully gapped
insulator.

  To construct excitations ``$\phi$'' (including intrinsic excitations and test particles), one must trivialize the mutual statistics between the excitation considered and $\phi_c$  such that the excitation is a deconfined particle which is observable in the excitation spectrum. According to Eq.(\ref{trivialm2}), the mutual statistics between $\phi_c$ and an excitation $\phi$ (labeled by quantum numbers without subscript ``$c$'') is trivialized by the following formula: 
\begin{align} 
N^{(1)}_ml+(N_M-N^{(1)}_m)l=s n^{(1)}_f-s n^{(2)}_f\,\label{constraintmutual}
\end{align} 
which constrains the quantum numbers of excitations leading to three independent labels instead of four. 
It may be equivalently expressed as:
\begin{align} 
&\ \ \ \ 
\frac{l}{s}N_M=n^{(1)}_f-n^{(2)}_f=  N^{(1)}_f-N^{(2)}_f\equiv  N^a
\,,\label{constraint}
\end{align} 
where the definition Eq.(\ref{na}) is applied, and, the condition (\ref{s0}) is implicit.    {Therefore, all excitations in the present mean-field ansatz can be uniquely labeled by $(N_M,N_m^{(1)},n^{(2)}_f)$ in the \emph{Standard Labeling}, while, $n^{(1)}_f$ is determined by Eq.(\ref{constraint}).}

%
%
%

Meanwhile, due to the screening effect shown in Eq.(\ref{aAaA}), the {\it total EM electric charge} $N_E$ is sum of $N_A$ (which is equal to $\alpha N^{(1)}_f+(1-\alpha) N^{(2)}_f$ according to Eq.(\ref{na})) and an additional screening part:
\begin{align}
N_E&=N_A-\frac{l}{s}N_m^a\,.\label{eq:totalcharge}
\end{align}
In the \emph{Standard Labeling}, $N_E$ is expressed as (details of derivation are present in Appendix \ref{appendix_eq:totalcharge}):
\begin{align}
N_E =& n^{(2)}_f- \frac{l}{s}N^{(1)}_m+2\alpha \frac{l}{s} N_M\,.   \label{chargestandard}
\end{align}

In fact, the condensed dyon has a trivial mutual statistics with itself. Thus, the total EM electric charge of the condensed dyon can also be calculated via Eq.(\ref{eq:totalcharge}):
\begin{align}
(N_E)_c=(N_A)_c-\frac{l}{s}(N^a_m)_c=l-s\frac{l}{s}=l-l=0
\end{align}
which indicates that the condensation indeed does not carry total EM electric charge and U(1) symmetry is exactly unbroken.

Under time-reversal symmetry transformation, $N_E$ has the following property:
\begin{align}
\overline{N_E}\equiv&\overline{n^{(2)}_f}-\frac{l}{s}\overline{N_m^{(1)}}+2\overline{\alpha}\frac{l}{s}\overline{N_M}=N_E\,,\label{eq:tot_charge}
\end{align}
where, Eq.(\ref{number2}) and Eq.(\ref{change1}) are applied, and $\overline{\alpha}=1-\alpha$ due to the exchange of $f_1$ and $f_2$. Since $(N_M,n^{(2)}_f,N^{(1)}_m)$ is an excitation (and thus (${N_E},N_M$) is a site on the charge lattice), we can prove that $(\overline{N_M},\overline{n^{(2)}_f},\overline{N^{(1)}_m})$ is also an excitation (and thus $(\overline{N_E},-N_M)$ is a site on the charge lattice) by justifying that $(\overline{N_M},\overline{n^{(2)}_f},\overline{N^{(1)}_m})$ satisfies the trivial mutual statistics condition Eq.(\ref{constraint}). Required by the time-reversal invariant mean-field ansatz we considered, the two dyons have the same excitation energy, such that Eq.(\ref{eq:tot_charge}) indicates that both the excitation energy and site distribution are mirror-symmetric at arbitrary $\alpha$.

 We have selected a time-reversal invariant dyon condensate $\phi_c$ but will the excitation spectrum (\emph{i.e.} charge lattice) respect time-reversal symmetry? According to Sec.\ref{sec:definingwitten}, in order to preserve  time-reversal symmetry, one must also  require that the charge lattice is mirror-symmetric about x-axis (including site distribution, quantum statistics, and excitation energy). As we have proved that site distribution and excitation energy are already mirror-symmetric shown in Eq.(\ref{eq:tot_charge}), the subsequent task is to examine whether quantum statistics are mirror-symmetric. 
 
 Generically, we expect that  only a sequence of $\alpha$ is allowed.  $\alpha=1/2$ satisfies Condition-\uppercase\expandafter{\romannumeral3}. In the remaining discussion of Sec.\ref{sec:82}, we will only focus on $\alpha=1/2$ which is the simplest choice in every mean-field ansatz. We will leave the discussion on the general $\alpha$-sequence 
to Sec.\ref{sec:general}.

Now we turn to the discussion of quantum statistics of excitations.  According to Eq.(\ref{quantumsign}), the statistics sign in the present mean-field ansatz ($\theta_1=\theta_2=0$, $\alpha=1/2$) can be obtained (details of derivation are present in Appendix \ref{appendix_eq:excitation_sign}):
\begin{align}
\Gamma\doteq~N_MN_E+(2N^{(1)}_m-N_M+1) \frac{l}{s}N_M\,,\label{eq:excitation_sign}
\end{align}
where, $N_E$ can be expressed as:
\begin{align}
N_E=n^{(2)}_f+\frac{l}{s}(N_M-N^{(1)}_m)\,\label{eq_fractional}
\end{align} 
by plugging $\alpha=1/2$ into Eq.(\ref{chargestandard}). In Eq.(\ref{eq:excitation_sign}), $N_E$ is explicitly written in order to compare $\Gamma$ with trivial Witten effect and non-trivial Witten effect  defined in Sec.\ref{sec:definingwitten}. One may also replace $N_E$ in Eq.(\ref{eq:excitation_sign})  by Eq.(\ref{eq_fractional}), rendering an equivalent expression of Eq.(\ref{eq:excitation_sign}):
\begin{align}
\Gamma\doteq~N_M\left(n^{(2)}_f+\frac{l}{s}(N^{(1)}_m+1)\right)\,.\label{eqabc}
\end{align}

 \begin{table*}
 \begin{tabular}[t]{|c|c|c|c|}
 \hline
 \begin{minipage}[t]{0.7in} \textbf{mean-field ansatz}\\ $(\theta_1,\theta_2)$\end{minipage} &\begin{minipage}[t]{1.4in}\textbf{trivial SPT\\(\emph{trivial Mott insulator of bosons})}\end{minipage}& \begin{minipage}[t]{1.5in}\textbf{non-trivial SPT\\(\emph{BTI: bosonic topological insulator})}\end{minipage}&\begin{minipage}[t]{3in}\textbf{SET \\ (\emph{fBTI: fractional bosonic topological insulator}) }\end{minipage}\\\hline\hline
     \begin{minipage}[t]{0.6in}~\\ $(0,0)$ \end{minipage}&\begin{minipage}[t]{1.4in}~\\ $\{l/s\in\mathbb{Z},l\neq 0\}\cup\{l=0,s=\pm 1 \}$\\\end{minipage} &\begin{minipage}[t]{1.5in} ~\\None\end{minipage} &\begin{minipage}[t]{3in} $\{l/s\notin\mathbb{Z}\}$;\\ Especially, $\{l=0, |s|\geqslant 2\}$ is a pure Z$_{|s|}$ TO state.\end{minipage}\\
\hline
  \begin{minipage}[t]{0.6in}~\\$(\pi,\pi)$\end{minipage}& \begin{minipage}[t]{1.4in}~\\ $\{{l}/{s} \in\mathbb{Z}\}$\end{minipage}& \begin{minipage}[t]{1in} ~\\None \end{minipage}&\begin{minipage}[t]{3in}~\\  $\{l/s\notin\mathbb{Z}\}$ (if $l/s=-1/2$, an additional Z$_{|s|}$ TO emerges.) \end{minipage}\\
    \hline\hline
  \end{tabular}
  \caption{Topological phases of bosons with U(1)$\rtimes$Z$^T_2$ symmetry in
three dimensions [labeled by $(\th_1,\th_2,\alpha=1/2,l,s)$].
All symmetric dyons are labeled by two integers $(l,s)$.
In each mean-field ansatz, different kinds of dyon condensations lead to,
generally, different topological phases (trivial SPT, non-trivial SPT, or,
SET). If $s=l=0$, \emph{i.e.} there is no dyon condensation, the resultant
symmetric state is the algebraic bosonic insulator (ABI) state which is gapless
and can be viewed as a parent state of all topological phases before condensing
some dyons. If $s=0,l\neq0$, the condensed dyon will break U(1) symmetry,
rendering a symmetry-breaking phase.  To get symmetric gapped phases, $s\neq 0$
has been required in the Table. The physical interpretation of $s$ and $l$ is
the following. In the mean-field ansatz $(\theta_1,\theta_2)=(0,0)$, the
condensed dyon with U(1)$\rtimes$Z$^T_2$ symmetry is a composite of $s$
monopoles of internal gauge field and $l$ physical bosons. In the mean-field
ansatz $(\theta_1,\theta_2)=(\pi,\pi)$, the condensed dyon with
U(1)$\rtimes$Z$^T_2$ symmetry is a composite of $s$ monopoles of internal gauge
field, $l$ physical bosons, and $s$ $f_2$ fermions.
    One ``physical boson'' is equal to one $f_1$ fermion plus one $f_2$
fermion.  ``Z$_{|s|}$ TO'' denotes the TO of Z$_{|s|}$ gauge theory which
arises from the gauge sector of the ground state. 
``None'' in a given entry means that the topological phase doesn't exist in the
corresponding mean-field ansatz. All trivial SPT has Witten effect with
$\Theta=0\,\,\text{mod}(4\pi)$ and all non-trivial SPT has Witten effect with
$\Theta=2\pi\,\,\text{mod}(4\pi)$. The discussion on Witten effect of SET will
be presented in Sec.\ref{sec:general} where trivial \emph{f}BTI and non-trivial
\emph{f}BTI are defined and classified. The mean-field ansatz $(0,\pi)$ always
breaks time-revesal symmetry.  
} 
\label{tab:results} 
\end{table*}

  \subsubsection{Different topological phases via different condensed dyons}\label{sec:00non}

\fbox{Bosonic intrinsic excitations}. With the above preparation, let us  study the nature of the U(1)$\rtimes$Z$^T_2$-symmetric topological (gapped) phases constructed via the condensed dyon $\phi_c$, mainly based on  Eqs.(\ref{constraint},\ref{eq:totalcharge},\ref{eq:excitation_sign},\ref{eq_fractional}). 
From Eq.(\ref{eq:excitation_sign}), all intrinsic excitations (carrying zero $N_M$) are bosonic, which \emph{rules out all fermionic intrinsic excitations} in the underlying boson system.   We also note that the $f_s$ fermions all have non-trivial
``mutual statistics'' with the  $\phi_c$ dyon, and thus those fermionic
excitations are confined.   Up to now, the only requirement on topological phases with symmetry is $s\neq 0$. To understand different topological phases via different condensed dyons, one needs to study the physical properties (quantum statistics, total EM electric charge) of all possible excitations constrained by Eq.(\ref{constraint}).  

\fbox{\{$l/s\in\mathbb{Z}$\}}. Let us  first focus on 
the parameter regime defined by $\frac{l}{s}\in\mathbb{Z}.$ In this case, Eq.(\ref{constraint}) allows excitations carrying arbitrary integer  $N_M$ and arbitrary integer $N_m^{(1)}$. In other words, an arbitrarily given integer $N_M$ can ensure that $N^a$ in right-hand-side of Eq.(\ref{constraint}) is integer-valued required by Eq.(\ref{alwaysinteger1}). The other two quantum numbers $N^{(1)}_m, n^{(2)}_f$ are still unconstrained and thus can take arbitrary integer and $N_E$ can also take arbitrary integer due to Eq.(\ref{eq_fractional}). To conclude, if $l/s\in\mathbb{Z}$, for any given integer combination ($N_M,N^{(1)}_m,n^{(2)}_f$), there exists at least one excitation. Thus, for any given integer combination ($N_E,N_M$), there exists at least one excitation on the charge lattice.   A useful corollary is that  for all intrinsic excitations ($N_M=0$), $N_E$ is always  integer-valued, which  {rules out intrinsic excitations carrying fractional EM electric charge (namely, ``\emph{fractional intrinsic excitation}'')}.

Then we will look for the general solutions of $(l,s)$ which admit trivial or non-trivial Witten effect. 
To look for the general solutions  which admit $\Gamma_{2}$, we solve the equation $\Gamma-\Gamma_{2}\doteq 0$, {\it i.e.}
\begin{align}
(2N^{(1)}_m-N_M+1)\frac{l}{s}N_M-N_M\doteq 0\,,\label{eq:2pi}
\end{align}
where, $N^{(1)}_m$ and $N_M$ are arbitrary integers if $l/s\in\mathbb{Z}$. 
 More precisely, if $N_M$ is an arbitrary even integer, we require that $N_M\cdot \frac{l}{s} (2N^{(1)}_m-N_M+1)$ is even integer. In other words, $\frac{l}{s} (2N^{(1)}_m-N_M+1)$ must be an integer for arbitrary integer $N_m^{(1)}$ and even integer $N_M$, which renders a requirement for the solution: $\frac{l}{s}$ must be integer-valued. This requirement is not new and is nothing but our starting point. On the other hand, if $N_M$ is an arbitrary odd integer, {\it i.e.} $N_M=2k+1$ where $k$ is an arbitrary integer, $\frac{l}{s}(2N^{(1)}_m-N_M+1)$ must be odd. In other words, $\frac{l}{s}(2N^{(1)}_m-2k)= \frac{2l}{s}(N^{(1)}_m-1)$ must be odd for arbitrary integer $N^{(1)}_m$. However, it is obviously even. Therefore, there is no solution admitting non-trivial SPT state. 

Likewise, to look for a solution which gives $\Gamma_0$, we solve the equation $\Gamma-\Gamma_0\doteq0$. It is easily obtained that the requirement is $\frac{l}{s}$ is an integer which is nothing but our starting point here. Therefore, if the two integers $l$ and $s$ satisfy that $\frac{l}{s}$ is an integer, such a choice $(l,s)$ is a solution which admits a state with a \emph{trivial Witten effect}. The state is a trivial SPT state by definition if TO doesn't exist.   One can check that the quantum statistics and sites are mirror-symmetric about x-axis. In addition, the distribution of excitation energy  is also mirror-symmetric due to Eq.(\ref{eq:tot_charge}). Overall, the charge lattice is indeed mirror-symmetric and thus satisfies Condition-\uppercase\expandafter{\romannumeral2}.

In fact we can also directly derive the trivial Witten effect result by reformulating $\Gamma$ in Eq.(\ref{eq:excitation_sign}) to $\Gamma\doteq~N_MN_E+(2N^{(1)}_m-N_M+1) \frac{l}{s}N_M\doteq N_MN_E$ where the last term in $\Gamma$ is always even once $l/s\in\mathbb{Z}$.  
 
The trivial Witten effect only rules out TO with fractional intrinsic excitations and TO with fermionic intrinsic excitations. Other TO patterns are still possible. Meanwhile, we note that actually the excitations can also come from pure gauge sector, in addition to matter field sector (\emph{i.e.} dyons) considered above. If $l=0,|s|\geqslant2, \frac{l}{s}=0$, the internal gauge symmetry U(1) of \emph{dynamical} gauge field $a_\mu$ is not fully broken but broken down to Z$_{|s|}$ gauge symmetry according to Eq.(\ref{Lagrangian}), which renders Z$_{|s|}$ TO in three-dimensions in the presence of global symmetry U(1)$\rtimes$Z$^T_2$. The low-energy field theory of this TO pattern is the \emph{topological BF theory} of level-$s$. The ground-state-degeneracy (GSD) on a three-torus is $|s|^3$.  Therefore, to get a trivial SPT state which doesn't admit any TO by definition,  one should further restrict the two integers $(l,s)$ satisfying: $\{\frac{l}{s}\in\mathbb{Z},l\neq0,s\neq0\}\cup \{l=0,s=\pm 1\}$.

 \fbox{\{$l/s\notin\mathbb{Z}$\}}.  If $l/s\notin\mathbb{Z}$, by noting that $N^a$ must  be  integer-valued in the present mean-field ansatz $\theta_1=\theta_2=0$, Eq.(\ref{constraint}) shows that the allowed EM magnetic charge $N_M$ of excitations cannot take arbitrary integer.  A direct example is that all particles with $N_M=1$ must be permanently confined since Eq.(\ref{constraint}) cannot be satisfied.  Despite that, the other two independent quantum numbers of excitations $N^{(1)}_f,N^{(1)}_m$ can still take arbitrary integer.

 For instance, if $l=1,s=3$, allowed value of $N_M$ should take $N_M=3k$ with $k\in\mathbb{Z}$, {\it i.e.} $N_M=0,\pm3,\pm6,\pm9,...$ in order to ensure the right-hand-side of Eq.(\ref{constraint}) is integer-valued. This quantization sequence is different from the sequence ($0,\pm1,\pm2,\pm3,...$) we are familiar with in the vacuum. By recovering full units (each boson carries a fundamental charge unit $e$), the EM magnetic charge $\frac{h}{e}\cdot 3k$ can be reexpressed as $\frac{h}{e^*}k$, where, $h$ is Planck constant, and, the effective fundamental EM electric charge  unit $e^*$ of intrinsic excitations is fractional: ``$e^*\equiv \frac{e}{3}$''. This fractional fundamental EM electric charge unit implies that the U(1)$\rtimes$Z$^T_2$-symmetric ground state constructed via condensing the dyon $\phi_c$  labeled by $(l,s)=(1,3)$ in the mean-field ansatz $\theta_1=\theta_2=0$ admits \emph{fractional intrinsic excitations} (which carry fractional EM electric charge), a typical signature of TO.  Interestingly, from Eq.(\ref{eq_fractional}), we find that  $N_E$ of excitations with $N_M=0$ indeed can take a fractional value. Therefore, the dyon excitations have self-consistently included fractional intrinsic excitations  in response to the new quantization sequence of the EM magnetic charge $N_M$.  In this sense, the topological phase labeled by $(l,s)=(1,3)$ contains TO (emergence of fractional intrinsic excitations) with global symmetry, \emph{i.e.} a SET state. 

Generally, we may parametrize $l/s=k'+\frac{p}{q}$, where, $k',p,q\in\mathbb{Z}$, $q>p>0$, $\text{gcd}(p,q)=1$ (gcd: greatest common divisor). Allowed excitations constrained by Eq.(\ref{constraint}) enforce  $N_M$ is quantized as  
\begin{align}
N_M=qk\,,\label{quantizationnm}
\end{align}
where, $k\in\mathbb{Z}$, \emph{i.e.} $N_M=0,\pm q,\pm2q,\pm3q,...$. Thus, the effective fundamental EM electric charge unit $e^*$ of intrinsic excitations should be consistently fractional, \emph{i.e.} $e^*=\frac{1}{q}\,e$. On the other hand, Eq.(\ref{eq_fractional}) shows that $N_E$ of all excitations may be fractional as a multiple of $1/q$:
\begin{align}
N_E=k_1 -pk_2/q\,,\label{totalEMcharge}
\end{align}
where,  the two integer variables $k_1,k_2$ are introduced
\begin{align}
 k_1\equiv n^{(2)}_f+k'(qk-N^{(1)}_m)+pk\,,\,
 k_2\equiv N^{(1)}_m\,\label{transitionlabel}
 \end{align}
 to replace $N^{(1)}_f$ and $N_m^{(1)}$. The first term in (\ref{totalEMcharge}) is always integer-valued, but the second term may be fractional as a multiple of $1/q$ by noting that $N^{(1)}_m$ is an arbitrary integer and $q>p>0$.  By setting $k=0$, we find that 
$N_E$ may be fractional with unit $e^*=1/q$.
In short, the state constructed here must be a SET state with fractional intrinsic excitations.

\begin{figure}[t]
\centering
\includegraphics[width=8.5cm]{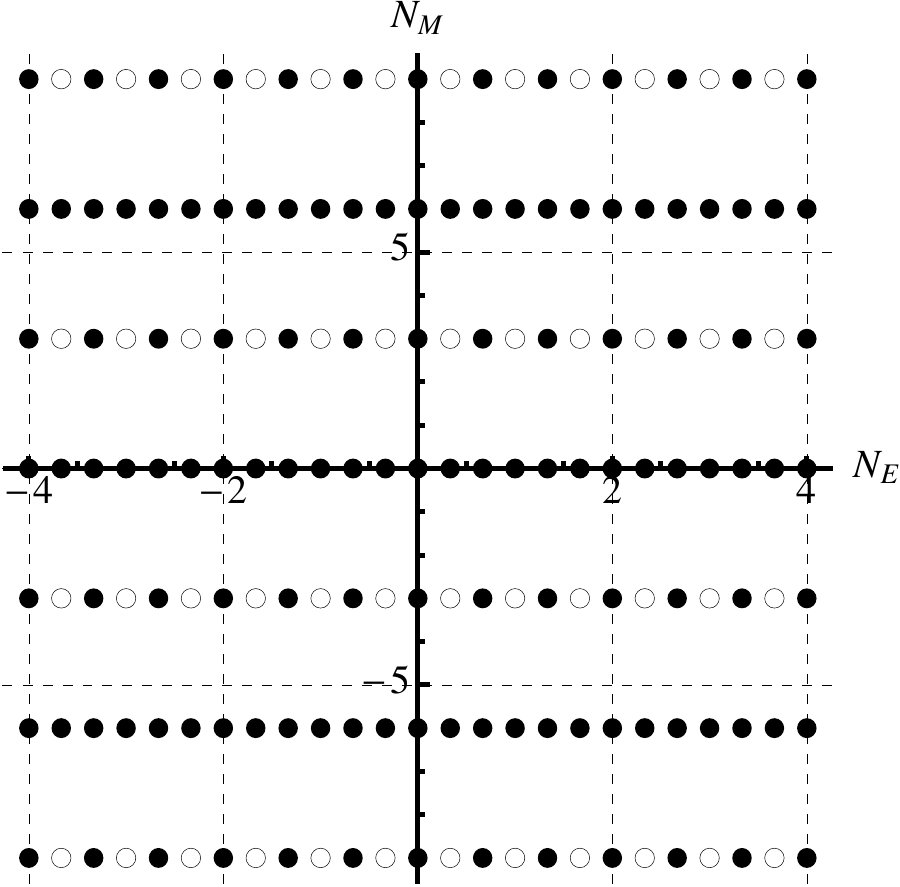}
\caption{Illustration of the Charge-Lattice-I (any $l/s$ satisfying $l/s=k'+1/3$, $k'\in\mathbb{Z}$) with fractional intrinsic excitations (in unit of $1/3$).  Excitations with $-4\leq N_E\leq 4$ and $-9\leq N_M\leq 9$ are plotted. The solid circles (open circles) denote bosonic (fermionic) statistics. }
\label{figure_chargelattice1}
\end{figure}

 Next, we will focus on the quantum statistics of all excitations defined in Eq.(\ref{eq:excitation_sign}). In  Eq.(\ref{eq:excitation_sign}),  the term including $N^{(1)}_m$ can be removed since  $2N^{(1)}_m \frac{l}{s}N_M =2N^{(1)}_m (qk'+p)k 
\doteq 0$. Therefore, $\Gamma$ is uniquely determined by $N_E, N_M$ in the following formula up to the quantization conditions (\ref{quantizationnm}) and (\ref{totalEMcharge}):
\begin{align} 
\Gamma=N_MN_E-N_M(N_M-1)\frac{l}{s}\,.\label{newgamma}
\end{align}
In short, by giving three arbitrary integers $(k,k_1,k_2)$, one can determine $N_E$ and $N_M$ via Eq.(\ref{totalEMcharge}) and Eq.(\ref{quantizationnm}) respectively, and further determine the quantum statistics of the excitations labeled by $(k,k_1,k_2)$. One can plot the quantum statistics in the  {charge lattice} formed by discrete data $N_{M}$ and $N_E$. We call this firstly introduced charge lattice with TO as ``\textbf{Charge-Lattice-I}''. One can check that the quantum statistics and sites are mirror-symmetric about x-axis. In addition, the distribution of excitation energy   is also mirror-symmetric due to Eq.(\ref{eq:tot_charge}).  Overall, the charge lattice is indeed mirror-symmetric and thus satisfies Condition-\uppercase\expandafter{\romannumeral2}. For example, if $l/s=k'+1/3$, the Charge-Lattice-I is shown in Fig.\ref{figure_chargelattice1}.

In short, this new SET  labeled by $l/s=k'+p/q$ with $k',p,q\in\mathbb{Z},q>p>0,\text{gcd}(q,p)=1$ (coined ``\emph{Dyonic TO}'' in order to distinguish it from TO which arises from the gauge sector, \emph{e.g.} Z$_{|s|}$ TO) has the following key properties:
\begin{enumerate}
\item All excitations are uniquely labeled by three arbitrary integers $(k,k_1,k_2)$ which are related to the \emph{Standard Labeling} via Eq.(\ref{transitionlabel}).
\item The total EM electric charge $N_E$ of intrinsic excitations is fractional with unit $e^*=1/q$ given by $N_E=k_1-\frac{p}{q}k_2$. The allowed EM magnetic charge $N_M$ is quantized at $q$. 
\item The quantum statistics of excitations is uniquely determined by $N_M$ and $N_E$ (cf. Eq.(\ref{newgamma})).
\item All intrinsic excitations  are bosonic.
 \end{enumerate}

 Before closing the analysis of the present mean-field ansatz, a possible confusion should be clarified. The TI state of free fermions admits $\Theta=\pi$ Witten effect and thus $N_E=n+\frac{1}{2}N_M$ ($n$ is integer number of attached fermions) may be fractional. But, this fractional EM electric charge is due to the presence of external EM magnetic monopole. To diagnose the TO of the ground state, we should restrict our attention to ``intrinsic excitations'' which requires $N_M=0$ by definition, as we are doing in the present work. Therefore, indeed the total EM electric charge of any intrinsic excitation in the TI is non-fractional and no TO exist.

\fbox{In summary}, in the mean-field ansatz with $(\theta_1,\theta_2)=(0,0)$, all symmetric gapped phases (condensed dyons with symmetry) are labeled by two integers $(s,l)$. Physically, a condensed dyon with symmetry labeled here is a composite of $l$ physical bosons (formed by $l$ $f_1$ fermions and $l$ $f_2$ fermions) and $s$ unit magnetic monopoles of internal gauge field $a_\mu$.   
The state is a trivial SPT state, {\it i.e.} a trivial Mott insulator of bosons with U(1)$\rtimes$Z$^T_2$ symmetry if  the two integers $(l,s)$ satisfy $\{\frac{l}{s}\in\mathbb{Z},l\neq0,s\neq0\}\cup \{l=0,s=\pm 1\}$. The state is a SET state if the two integers $(l,s)$ either \emph{(i)} satisfy $\{l=0,|s|\geqslant 2\}$, which corresponds to the SET state with Z$_{|s|}$ TO, or \emph{(ii)} satisfy $\{l/s\notin\mathbb{Z}\}$ which corresponds to a new SET---``Dyonic TO''.

 We note that if the fermions $f_1$ and $f_2$ in our construction are replaced by two bosons both of which are in trivial Mott insulator states of bosons (\emph{i.e.} $\theta_1=\theta_2=0$ and $\alpha=1/2$), there exists a solution for non-trivial SPT state. For example, $l=s=1$ is a candidate of non-trivial SPT state, {\it i.e.} a non-trivial BTI state, which corresponds to condensation of one physical boson attached to a magnetic monopole of internal gauge field $a_\mu$.\cite{MM13a} The charge lattice is shown in Fig.\ref{figure_chargelattice2pi}. This can be verified easily with the same technique shown above. The only difference is that in this bosonic projective construction, the terms $\prod_s(-1)^{n^{(s)}_f}$ in the quantum statistics formula (\ref{quantumsign}) should be removed for the reason that attached particles are bosonic.

 \subsection{Mean-field ansatz $(\theta_1,\theta_2)=(\pi,\pi)$}

\subsubsection{Dyon condensation with symmetry}

Following the same strategy, we can also consider the mean-field ansatz with $(\theta_1,\theta_2)=(\pi,\pi)$ where both fermions are in non-trivial TI states. In this case,  Condition-\uppercase\expandafter{\romannumeral1} restricts $(N^{(1)}_f)_c=(N^{(2)}_f)_c$. Therefore, a general dyon with time-reversal symmetry is labeled by two integers $l$ and $s$: 
\begin{align}
(N^a_m)_c=s\,,(N^{(1)}_f)_c=l+\frac{s}{2}\,,(N^{(2)}_f)_c=l+\frac{s}{2}\,,(N_M)_c=0\,.
\end{align}
One can check that the dyon condensate is time-reversal symmetric and $(N^a_m)_c$ is quantized at integer with any given $\alpha$ in the present mean-field ansatz. And such a general dyon is always bosonic since $(\Gamma)_c=$even integer. Most importantly, by definition, the EM field here is a probe field such that by switching off EM field, the physical ground state should not carry EM magnetic charge. So, $(N_M)_c$ must be vanishing, which is also required by time-reversal symmetry. Other quantum numbers of the condensed dyon are straightforward: 
\begin{align}
&(N_A)_c=l+\frac{s}{2}\,,(N^{(1)}_m)_c=s\,,(N^{(2)}_m)_c=-s\,,\nonumber\\
&(N^a)_c=0\,,(n^{(1)}_f)_c=l\,,(n^{(2)}_f)_c=l+s.
\end{align}

 But will such a dyon condensed state respect the EM electric U(1) symmetry and behave like
an insulator? To answer this question, according to Sec.\ref{sec:time-reversal}, let us write down the effective Lagrangian of the
condensed dyon $\phi_c$ in real time (only spatial components are written here for simplicity and time component is similar):
\begin{align} 
 \mathscr{L}_{\rm kin}&=\frac{1}{2 m}| (-i\nabla+s\widetilde{\mathbf{a}}+(l+\frac{s}{2})\mathbf{A})\phi_c |^2
-V(\phi_c)\,,\label{Lagrangianpipi}
 \end{align}
where, $V(\phi_c)$ is a symmetric potential energy term which energetically stabilizes the bosonic condensate.

In the dyon condensed state $\phi_c\neq 0$,
the internal gauge field $a_\mu$ is gapped and satisfies 
\begin{align}
s\widetilde{\mathbf{a}}=-(l+\frac{s}{2}) \mathbf{A}\,\label{aAaA1}
\end{align} 
which indicates that the internal gauge field cannot fluctuate freely and is locked to the non-dynamical EM background.  We also require that 
\begin{align}
s\neq 0\,
\end{align}
in the following in order that the
dyon condensation does not generate the $\mathbf{A}^2$ term. 
This requirement may be understood in the following way. If $s=l=0$, there is no dyon condensation, which is nothing but the ABI state discussed in Sec.\ref{sec:abi}. If $s=0$ and $l\neq0$, Eq.(\ref{aAaA1}) reduces to $\mathbf{A}=0$ which is the consequence of mass term $\mathbf{A}^2$, a fingerprint of superconductor/superfluid with broken U(1). This case is nothing but condensation of $l$-bosons (carrying EM electric charge $l$), which breaks U(1) symmetry down to Z$_{|l|}$ discrete symmetry spontaneously ($Z_1$ represents complete breaking of U(1)).   
In the following, in order to preserve U(1) symmetry and consider dyon condensation, we will restrict our attention to $s\neq 0$. Thus the  dyon
condensed state indeed respects the EM electric U(1) symmetry and represents a fully gapped
insulator. 

It should be noted that the surface fermionic gapless excitations described
by two $f_s$ Dirac fermions are also confined by the  $\phi_c$
condensation.  The confinement behaves like a strong attraction between $f_1$
and $f_2$ fermions,  which may make
the surface into a superconducting state.

  To construct excitations ``$\phi$'' (including intrinsic excitations and test particles), one must trivialize the mutual statistics between the excitation $\phi$ and condensed dyon $\phi_c$  such that the excitation is a deconfined particle which is observable in the excitation spectrum. According to Eq.(\ref{trivialm2}), the mutual statistics between an excitation $\phi$ labeled by  quantum numbers without subscript ``$c$'' and $\phi_c$ is trivialized by the following formula: 
\begin{align} 
&N^{(1)}_ml+(N_M-N^{(1)}_m) (l+s) 
=sn^{(1)}_f-sn^{(2)}_f\,\label{constraintmutual1}
\end{align} 
which leads to:
\begin{align} 
 n^{(1)}_f-n^{(2)}_f=(\frac{l}{s}+1)N_M-N^{(1)}_m\,.\label{constraint1}
\end{align} 
Note that, by considering Eqs.(\ref{eq:W},\ref{na},\ref{constraint1}), we obtain an alternative expression of Eq.(\ref{constraint1}):
\begin{align} 
N^a=(\frac{l}{s}+\frac{1}{2})N_M\,.\label{constraint001}
\end{align}

Eq.(\ref{constraint1}) is an important and unique constraint on the quantum numbers of excitations $\phi$ constructed above the condensed dyon $\phi_c$. In other words, a dyon is served as a deconfined particle (\emph{i.e.} an excitation with a finite gap) above the condensed dyon $\phi_c$ if its four quantum numbers are constrained by Eq.(\ref{constraint1}).   {Therefore, all excitations in the present mean-field ansatz can be uniquely labeled by $(N_M,N_m^{(1)},n^{(2)}_f)$ in the \emph{Standard Labeling}, while, $n^{(1)}_f$ is determined by Eq.(\ref{constraint1}).}

Meanwhile, due to the screening effect shown in Eq.(\ref{aAaA1}), the {\it total EM electric charge} $N_E$ is sum of $N_A$ and a screening part:
\begin{align}
N_E=&N_A-\frac{l+\frac{s}{2}}{s}N_m^a\,.
\label{eq:totalcharge1}
\end{align}

In the \emph{Standard Labeling}, $N_E$ is expressed as (details of derivation are present in Appendix \ref{appendix_chargestandard1}):
\begin{align}
N_E =&-(\frac{l}{s}+1)N^{(1)}_m+n^{(2)}_f+(2\alpha \frac{l}{s}+\alpha+\frac{1}{2})N_M
\,.   \label{chargestandard1}
\end{align}

In fact, the condensed dyon has a trivial mutual statistics with itself. Thus, the total EM electric charge of the condensed dyon can also be calculated via Eq.(\ref{eq:totalcharge1}):
\begin{align}
(N_E)_c=&(N_A)_c-(N^a_m)_c\cdot \frac{l+\frac{s}{2}}{s}\,\nonumber\\
=&l+\frac{s}{2}-s(l+\frac{s}{2})/s=0
\end{align}
which indicates that the condensation indeed does not carry total EM electric charge such that U(1) symmetry is exactly unbroken. 

Under time-reversal symmetry transformation, $N_E$ has the following property:
\begin{align}
\overline{N_E}=&-(\frac{l}{s}+1)\overline{N^{(1)}_m}+\overline{n^{(2)}_f}+(2\overline{\alpha} \frac{l}{s}+\overline{\alpha}+\frac{1}{2})\overline{N_M}=N_E  \,. \label{eq:tot_chargepipi}
\end{align}
Since $(N_M,n^{(2)}_f,N^{(1)}_m)$ is an excitation (and thus (${N_E},N_M$) is a site on the charge lattice), we can prove that $(\overline{N_M},\overline{n^{(2)}_f},\overline{N^{(1)}_m})$ is also an excitation (and thus $(\overline{N_E},-N_M)$ is a site on the charge lattice) by justifying that $(\overline{N_M},\overline{n^{(2)}_f},\overline{N^{(1)}_m})$ satisfies the trivial mutual statistics condition Eq.(\ref{constraint}). Required by the time-reversal invariant mean-field ansatz we considered, the two dyons have the same excitation energy, such that Eq.(\ref{eq:tot_charge}) indicates that both the excitation energy and site distribution are mirror-symmetric at arbitrary $\alpha$.

Likewise, we choose the simplest case: $\alpha=1/2$ in this Section. Plugging $\alpha=1/2$ into Eq.(\ref{chargestandard1}), we obtain: 
\begin{align}
N_E=-(\frac{l}{s}+1)N^{(1)}_m+n^{(2)}_f+(\frac{l}{s}+1)N_M\,.\label{eq:tot_charge2}
\end{align}

Now we turn to the discussion of quantum statistics of excitations.  According to Eq.(\ref{quantumsign}), the statistics sign in the present mean-field ansatz ($\theta_1=\theta_2=\pi$, $\alpha=1/2$) is (details of the derivation are present in Appendix \ref{appendix_eq:excitation_sign1}):
\begin{align}
\Gamma\doteq~N_M(N^{(1)}_m+1)(\frac{l}{s}+1)+N_M n^{(2)}_f\,.
\label{eq:excitation_sign1}
\end{align}

\subsubsection{Different topological phases via different condensed dyons}

With the above preparation, let us  study the nature of the U(1)$\rtimes$Z$^T_2$-symmetric topological (gapped) phases constructed above the condensed dyon $\phi_c$, mainly based on Eqs.(\ref{constraint1},\ref{eq:tot_charge2},\ref{eq:excitation_sign1}).

\fbox{Bosonic intrinsic excitations}. From Eq.(\ref{eq:excitation_sign1}), all intrinsic excitations with zero $N_M$ are bosonic (excitations with $N_M=0$ always exist in Eq.(\ref{constraint1})), which rules out all fermionic intrinsic excitations in the underlying boson system.   We also note that the $f_s$ fermions all have non-trivial
``mutual statistics'' with the  $\phi_c$ dyon, and thus those fermionic
excitations are confined.   
 Up to now, the only requirement on topological phases with symmetry is $s\neq 0$. To understand different topological phases via different condensed dyons, one needs to study the physical properties (quantum statistics, total EM electric charge) of all possible excitations constrained by Eq.(\ref{constraint1}). 

\fbox{\{$l/s\in\mathbb{Z}$\}}. Let us  first focus on 
the parameter regime defined by $\frac{l}{s}\in\mathbb{Z}$. In this case, Eq.(\ref{constraint1}) allows excitations carrying arbitrary integers $N_M$, $N^{(1)}_m$, and $n^{(2)}_f$. $n^{(1)}_f$ is uniquely fixed by Eq.(\ref{constraint1}). And, $N_E$ in Eq.(\ref{eq:tot_charge2}) is also fixed and can also take arbitrary integer.  
   
  We may reformulate $\Gamma$ in Eq.(\ref{eq:excitation_sign1}) by means of $l/s\in\mathbb{Z}$ (details of derivation are present in Appendix \ref{appendix_eqa}):
\begin{align}
\Gamma\doteq N_M N_E\,.\label{eqa}
 \end{align}
 Due to Eq.(\ref{constraint1}), $n^{(2)}_f$ and $N^{(1)}_m$ can be still arbitrarily integer-valued, and, $n^{(1)}_f$ is fixed once $N_m^{(1)}$, $N_M$ and $n^{(2)}_f$ are given.   Thus, $N_E$ in Eq.(\ref{eq:tot_charge2}) can take any integer. There exists at least one excitation for any given integer combination $(N_E,N_M)$.  
In short, if the two integers $l$ and $s$ satisfy that $\frac{l}{s}$ is an integer, such a choice $(l,s)$ is a solution which admits a state with a \emph{trivial Witten effect}. The state is a trivial SPT state. One can check that the quantum statistics and sites are mirror-symmetric about x-axis. In addition, the distribution of excitation energy  is also mirror-symmetric due to Eq.(\ref{eq:tot_chargepipi}).    Overall, the charge lattice is indeed mirror-symmetric and thus satisfies Condition-\uppercase\expandafter{\romannumeral2}.

\fbox{$\{l/s\notin\mathbb{Z}$\}}. We note that $l/s=-1/2$ is a special point where the internal gauge symmetry U(1) is broken down to Z$_{|s|}$ gauge symmetry according to Eq.(\ref{Lagrangianpipi}). It leads to  Z$_{|s|}$ TO in three-dimensions in the presence of global symmetry U(1)$\rtimes$Z$^T_2$.  In the following, we will not consider this point. 

For a general parameter choice in $l/s\notin\mathbb{Z}$, we will see that there is a Dyonic TO which is defined as TO arising from dyons. Generally, we may parametrize $l/s=k'+\frac{p}{q}$, where, $k',p,q\in\mathbb{Z}$, $q>p>0$, $\text{gcd}(p,q)=1$ (gcd: greatest common divisor).  Plugging $l/s=k'+\frac{p}{q}$ into Eq.(\ref{constraint1}), we find that $N_M$ must be quantized at $q$ in all allowed excitations constrained by Eq.(\ref{constraint1}). That is, 
\begin{align}
N_M=qk\,,\label{quantizationnm1}
\end{align}
where, $k\in\mathbb{Z}$, \emph{i.e.} $N_M=0,\pm q,\pm2q,\pm3q,...$.  On the other hand, Eq.(\ref{eq:tot_charge2}) shows that $N_E$ of all excitations may be fractional as a multiple of $1/q$:
\begin{align}
N_E=k_1 -pk_2/q\,,\label{natot}
\end{align}
where,  the two integer variables $k_1,k_2$ are introduced and related to the \emph{Standard Labeling} in the following way: 
\begin{align}
&k_1\equiv (k'+1)qk+pk+n^{(2)}_f-(k'+1)N^{(1)}_m\,,\\
&k_2\equiv N^{(1)}_m\,.
\end{align}

Due to Eq.(\ref{constraint1}), $n^{(2)}_f$ and $N^{(1)}_m$ can be still arbitrarily integer-valued, and, $n^{(1)}_f$ is fixed once $N_m^{(1)}$, $N_M$ and $n^{(2)}_f$ are given.   Thus the new variables $k_1$ and $k_2$ can be any integers. Hereafter, all excitations are labeled by the three independent integers $(k,k_1,k_2)$. Using these new labels, we see that $N_E$ of intrinsic excitations ($k=0$) can be still fractional according to Eq.(\ref{natot}) which doesn't depends on $k$. It indicates that the state constructed here is a SET state with  fractional intrinsic excitations. A useful observation from Eq.(\ref{natot}) is that $N_E$ can also take any integer once $k_2=q$.

 Next, we will focus on the quantum statistics of all excitations defined in Eq.(\ref{eq:excitation_sign1}). In the present parameter regime, $\Gamma$ can be expressed as: (details of derivation are present in Appendix \ref{appendix_eqa1})
  \begin{align}
\Gamma\doteq N_MN_E-\frac{l}{s} N_M(N_M-1)\,. \label{eqa1}
 \end{align}
Therefore, $\Gamma$ is uniquely determined by $N_E, N_M$ in the following formula up to the quantization conditions (\ref{quantizationnm1}) and (\ref{natot}). In short, by giving three arbitrary integers $(k,k_1,k_2)$, one can determine $N_E$ and $N_M$ via Eq.(\ref{natot}) and Eq.(\ref{quantizationnm1}) respectively, and further determine the quantum statistics of the excitations labeled by $(k,k_1,k_2)$. One can plot the quantum statistics in the  {charge lattice} expanded by discrete variables $N_{M}$ and $N_E$, which is same as \textbf{Charge-Lattice-I} discussed in Sec.\ref{sec:00non} and shown in Fig.\ref{figure_chargelattice1} ($l/s=1/3$). One can check that the quantum statistics and sites are mirror-symmetric about x-axis. In addition, the distribution of excitation energy is also mirror-symmetric due to Eq.(\ref{eq:tot_chargepipi}).   Overall, the charge lattice is indeed mirror-symmetric and thus satisfies Condition-\uppercase\expandafter{\romannumeral2}.

     \fbox{In summary}, in the mean-field ansatz with $(\theta_1,\theta_2)=(\pi,\pi)$,  all symmetric gapped phases (condensed dyons with symmetry) are labeled by two integers $(s,l)$. Physically, a condensed dyon with symmetry labeled here is a composite of $l$ physical bosons (formed by $l$ $f_1$ fermions and $l$ $f_2$ fermions), $s$ $f_2$ fermions, and $s$ unit magnetic monopoles of internal gauge field $a_\mu$.    If ${l}/{s}\in\mathbb{Z}$, the ground state is a trivial SPT state, \emph{i.e.} trivial Mott insulator of bosons.   
     In the parameter regime $\{l/s\notin\mathbb{Z}, l/s \neq-1/2\}$, the ground state is a SET state with  Dyonic TO (fractional intrinsic excitations).  If $l/s=-1/2$, the ground state is a SET state with  Z$_{|s|}$ TO.

\section{Topological phases with symmetry: General $\alpha$-sequence}\label{sec:general}

\subsection{Main results}

In Sec. \ref{sec:82}, we have obtained many topological phases based on dyon
condensations (see Table \ref{tab:results}). The value of $\alpha$ is chosen to
be $\alpha=1/2$ such that both $f_1$ and $f_2$ carry $1/2$ EM electric charge
(see Table \ref{chargetable}).  Such a choice preseves the time-reversal
symmetry. If we choose $\al$ to be some other values, the time-reversal
symmetry may be broken.  However, $\alpha=1/2$ is not the only value that
potentially preserves  time-reversal symmetry.  In the following, we shall study the general
$\alpha$-sequence which respects time-reversal symmetry. Since  the mean-field
ansatz $(0,\pi)$ always breaks time-reversal symmetry, we only consider other
two ansatzes. The main results are summarized in Table \ref{tab:results1}. We
note that the results in Table \ref{tab:results} can be obtained by taking
$\alpha=1/2$ in Table \ref{tab:results1}.

\begin{table*}
 \begin{tabular}[t]{|c|c|c|c|}
 \hline
 \begin{minipage}[t]{0.67in} \textbf{mean-field ansatz}\\ $(\theta_1,\theta_2)$\end{minipage} &\begin{minipage}[t]{2.1in}\textbf{trivial SPT\\(\emph{trivial Mott insulator of bosons})}\end{minipage}& \begin{minipage}[t]{1.7in}\textbf{non-trivial SPT\\(\emph{BTI: bosonic topological insulator})}\end{minipage}&\begin{minipage}[t]{2.3in}\textbf{SET} \\ \textbf{(\emph{fBTI: fractional bosonic topological insulator})} \end{minipage}\\\hline\hline
     \begin{minipage}[t]{0.6in} ~\\ ~\\~\\ $(0,0)$\\ \end{minipage}&\begin{minipage}[t]{2in} ~\\ ~\\$\{l/s=\text{odd},\alpha =\text{half odd}\}\cup\{l/s=\text{even},l\neq0,2\alpha=\text{integer}\}\cup\{l=0,s=\pm 1 ,2\alpha=\text{integer}\}$\\\end{minipage} &\begin{minipage}[t]{1.7in} ~\\  ~\\ ~\\$\{l/s=\text{odd},\alpha=\text{integer}\}$\end{minipage} &\begin{minipage}[t]{2.3in}  $\{l=0, |s|\geqslant 2,2\alpha=\text{integer}\}$: Z$_{|s|}$ TO and $\Theta=0\,\,\text{mod}\,{4\pi}$\\~\\ \{$l/s\notin\mathbb{Z}$\}, \emph{e.g.}  ${l}/{s}=\frac{1}{3}$: \\$\alpha=\text{half-odd}$ $\longrightarrow$ trivial SET with $\Theta=0\,\,\text{mod}\,\frac{4\pi}{9}$; \\ $\alpha=\text{integer}$ $\longrightarrow$ non-trivial SET with $\Theta=\frac{2\pi}{9} \,\text{mod} \frac{4\pi}{9}$ Witten effect\end{minipage}\\
\hline
  \begin{minipage}[t]{0.6in} ~\\ ~\\ ~\\ $(\pi,\pi)$ \end{minipage}& \begin{minipage}[t]{2in}~\\  ~\\~\\$\{l/s\in\mathbb{Z},\alpha-\frac{1}{2}=\text{even}\}$\\  \end{minipage}& \begin{minipage}[t]{1.7in} ~\\ ~\\ ~\\$\{l/s\in\mathbb{Z},\alpha-\frac{1}{2}=\text{odd}\}$\\ \end{minipage}&\begin{minipage}[t]{2.3in} $\{l/s=-1/2, \alpha=\text{half-odd}\}$: Z$_{|s|}$ TO and $\Theta=\frac{\pi}{2} \,\text{mod}\, \pi$ Witten effect\\~\\  \{$l/s\notin\mathbb{Z},l/s\neq-\frac{1}{2}$\}, \emph{e.g.}  ${l}/{s}=\frac{1}{3}$: \\$\alpha-\frac{1}{2}=\text{even}$ $\longrightarrow$ trivial \emph{f}BTI with $\Theta=0\,\,\text{mod}\,\frac{4\pi}{9}$; \\ $\alpha-\frac{1}{2}=\text{odd}$ $\longrightarrow$ non-trivial \emph{f}BTI with $\Theta=\frac{2\pi}{9} \,\text{mod} \,\frac{4\pi}{9}$ Witten effect\end{minipage}\\
    \hline\hline
  \end{tabular}
  \caption{Topological phases of bosons with U(1)$\rtimes$Z$^T_2$ symmetry in
three dimensions (for a generic $\alpha$-sequence), labeled by $(\th_1,\th_2,\alpha,l,s)$. A state is U(1)$\rtimes$Z$^T_2$ symmetric if the parameters $l,s, \alpha$ satisfy the conditions in
this table. 
We see that if $\alpha=1/2$, all allowed SPT states are trivial in any
mean-field ansatz in consistent to Table \ref{tab:results}.  All SET
states can be further classified into trivial \emph{f}BTI (without Witten
effect) and non-trivial \emph{f}BTI (admitting Witten effect). The mean-field
ansatz $(0,\pi)$ always breaks time-reversal symmetry.  }
  \label{tab:results1}
\end{table*}

\subsection{Mean-field ansatz   $(\theta_1,\theta_2)=(0,0)$}\label{sec:00alpha}

In this mean-field ansatz, according to the general statement in Sec.\ref{axionfield}, two allowed values of $\alpha$ must be differed from each other by any integer, which is required by charge quantization argument. Let us consider $\overline{\alpha}=1-\alpha$ and $\alpha$ where $\overline{\alpha}$ is the time-reversal transformed $\alpha$ shown in Eq.(\ref{alphabar}). The requirement $\overline{\alpha}-\alpha=\text{any integer}$ is equivalent to the constraint $2\alpha=\text{integer}$ which is nothing but Condition-\uppercase\expandafter{\romannumeral3}. This is the first constraint we obtained on the domain value of $\alpha$. 

We have obtained the total EM electric charge $N_E$ in the \emph{Standard Labeling} and in the presence of $\alpha$ (cf. Eq.(\ref{chargestandard})). To approach a mirror-symmetric charge lattice, we  require that the site distribution, excitation energy and quantum statistics are mirror-symmetric (cf. Condition-\uppercase\expandafter{\romannumeral2}).

\fbox{$l/s\in\mathbb{Z}$}. Let us first consider $l/s\in\mathbb{Z}$ such that $N_M$ is arbitrarily integer-valued due to the constraint Eq.(\ref{constraint}). We assume that the mirror site of $(N_M,n_f^{(2)}, N^{(1)}_m)$ is labeled by $(-N_M, n_f^{(2)'}, N^{(1)'}_m)$. In order that the mirror site does exist on the charge lattice, the integer solutions $(n_f^{(2)'}, N^{(1)'}_m)$ of the following equation must exist for any given integer $N_M$ (see Eq.(\ref{chargestandard})):
 \begin{align}
N_E =& n^{(2)'}_f- \frac{l}{s}N^{(1)'}_m-2\alpha \frac{l}{s} N_M\,
\end{align}
which is equivalent to
 \begin{align}
\left(n^{(2)'}_f-n^{(2)}_f\right)- \frac{l}{s}\left(N^{(1)'}_m- N^{(1)}_m\right)=4\alpha \frac{l}{s} N_M\,\label{nm1}
\end{align}
by means of Eq.(\ref{chargestandard}).    Therefore, the mirror-symmetric site distribution requires that:  $4\alpha\frac{l}{s}=\text{integer}$. To construct symmetric topological phases, we need further check the quantum statistics in the presence of $\alpha$ (details of derivation are present in Appendix \ref{appendix_q1}):
\begin{align}
\Gamma=&N_M \left(N_E-2\alpha\frac{l}{s}N_M+\frac{l}{s}\right)\,.\label{q1}
\end{align}
Therefore, the quantum statistics is mirror-symmetric if
\begin{align}
&N_M \left(N_E-2\alpha\frac{l}{s}N_M+\frac{l}{s}\right)\nonumber\\
\doteq&(-N_M )\left(N_E-2\alpha\frac{l}{s}(-N_M)+\frac{l}{s}\right)\nonumber\\
&\text{\emph{i.e.} }2N_M N_E\doteq 0\,.\label{alphacondition}
\end{align}
In other words, $2N_M N_E$ must be always even.

If $4\alpha\frac{l}{s}=\text{odd}$, Eq.(\ref{chargestandard}) indicates that $N_E$ is half-odd integer if we take $N_M=1$. As a result, Eq.(\ref{alphacondition}) is not satisfied.   Therefore, in order to guarantee mirror-symmetric distribution of quantum statistics, we need to consider a stronger condition:  $4\alpha\frac{l}{s}=\text{even}$, \emph{i.e.} $2\alpha\frac{l}{s}=\text{integer}$.  In Eq.(\ref{eq:tot_charge}), we have already proved that energy is mirror-symmetric for any $\alpha$, so that we conclude that to obtain a mirror-symmetric charge lattice (\emph{i.e.} Condition-\uppercase\expandafter{\romannumeral2}), we need $2\alpha\frac{l}{s}=\text{integer}$.  
Under this condition as well as $2\alpha=\text{integer}$, we may obtain   trivial SPT states and non-trivial SPT states summarized in Table \ref{tab:results1} by comparing $\Gamma$ with the standard trivial Witten effect and non-trivial Witten effect defined in Sec.\ref{sec:definingwitten}. As usual, one should pay attention to the emergence of Z$_{|s|}$ TO if $l=0$ and $|s|\geqslant 2$ although charge lattice formed by deconfined dyons is same as a trivial SPT state.  Strikingly, we obtain BTI states which are completely absent in Table \ref{tab:results} where $\alpha=1/2$ is fixed.

Since $N_M\in\mathbb{Z}$, $\frac{l}{s}\in\mathbb{Z}$ and $\frac{l}{s}N_M\doteq\frac{l}{s}(N_M)^2$, one may rewrite Eq.(\ref{q1}) as $\Gamma\doteq N_M(N_E-2\alpha\frac{l}{s}N_M+\frac{l}{s}N_M)\equiv N_M(N_E-\frac{\Theta}{2\pi}N_M)$. The minimal periodicity of $\Theta$ is $4\pi$ because $\Gamma$ is invariant after $4\pi$ shift. As a result, a $\Theta$ angle can be formally defined:
\begin{align}
{\Theta}\equiv-2\pi\frac{l}{s}+4\pi\frac{l}{s}\alpha\,\,\,\text{mod} (4\pi)\,\label{Theta00}
\end{align}
from which we see that the $\Theta$ angle is linearly related to $\alpha$.

\fbox{$l/s\notin\mathbb{Z}$}. Generally, we may parametrize $l/s=k'+\frac{p}{q}$, where, $k',p,q\in\mathbb{Z}$, $q>p>0$, $\text{gcd}(p,q)=1$ (gcd: greatest common divisor). In this case, $N_M$ is quantized at $qk$ as shown in Eq.(\ref{quantizationnm}).  To guarantee mirror-symmetric site distribution, the integer solutions $(n_f^{(2)'}, N^{(1)'}_m)$ of Eq.(\ref{nm1}) must exist for any given  $N_M=qk$:
 \begin{align}
&\left(n^{(2)'}_f-n^{(2)}_f\right)- (k'+\frac{p}{q})\left(N^{(1)'}_m- N^{(1)}_m\right)\nonumber\\
=&4\alpha(qk'+p)k\,\label{nm2}
\end{align}
by means of Eq.(\ref{chargestandard}).  Eq.(\ref{q1}) is also valid when $l/s\notin\mathbb{Z}$ by noting that $-2N_M\frac{l}{s}N_m^{(1)}$ is still even integer in deriving the fourth line of Appendix \ref{appendix_q1}.  Therefore, Eq.(\ref{alphacondition}) is also valid when $l/s\notin\mathbb{Z}$.

A general discussion on Eq.(\ref{nm2}) and Eq.(\ref{alphacondition}) is intricate. Let us take a simple example: $l/s=1/3$, \emph{i.e.} $k'=0,q=3,p=1$. The right hand side of Eq.(\ref{nm2}) becomes $4\alpha k$. To obtain the integer solutions $(n_f^{(2)'}, N^{(1)'}_m)$ for any given  integers  ($k, N_m^{(1)}, n^{(2)}_f$),  a constraint on $\alpha$ is necessary: $\alpha={\text{integer}}/{12}$. Under this condition, Eq.(\ref{alphacondition}) leads to a stronger condition: $6\alpha=k_0$ where $k_0$ is an integer. It  guarantees mirror-symmetric distribution of both sites and quantum statistics. As we have proved, energy is already mirror-symmetric due to Eq.(\ref{eq:tot_charge}). Overall, to obtain a mirror-symmetric charge lattice (\emph{i.e.}Condition-\uppercase\expandafter{\romannumeral2}), we need $\alpha=\text{integer}/6$. Keeping in mind that $2\alpha=\text{integer}$ required by Condition-\uppercase\expandafter{\romannumeral3}, the two conditions altogether still give $2\alpha=\text{integer}$. 

Since $N_M/3\in\mathbb{Z}$, $\frac{l}{s}=1/3$ and $\frac{l}{s}N_M=N_M/3\doteq (N_M)^2/9$, one may rewrite Eq.(\ref{q1}) as $\Gamma\doteq N_M(N_E-\frac{2\alpha}{3}N_M+\frac{1}{9}N_M)\equiv N_M(N_E-\frac{\Theta}{2\pi}N_M)$. The minimal periodicity of $\Theta$ is $\frac{4\pi}{9}$ because $\Gamma$ is invariant after $\frac{4\pi}{9}$ shift. As a result, a $\Theta$ angle can be formally defined:
\begin{align}
{\Theta}\equiv -\frac{2\pi}{9} +\frac{4\pi}{3}\alpha\,\,\,\text{mod} (\frac{4\pi}{9})\,\label{Theta00fractional}
\end{align}
from which we see that the $\Theta$ angle is linearly related to $\alpha$.   All SET states have fractional intrinsic excitations.  We can further classify these states into two categories: one is $\Theta=0\,\,\,\text{mod} (\frac{4\pi}{9})$ with $\alpha=\text{half-odd}$; one is $\Theta=\frac{2\pi}{9}\,\,\,\text{mod} (\frac{4\pi}{9})$ with $\alpha=\text{integer}$. In comparison to the trivial and non-trivial SPT states, we call the former ``trivial \emph{f}BTI" and the latter ``non-trivial \emph{f}BTI" via investigating the Witten effect.

\subsection{Mean-field ansatz   $(\theta_1,\theta_2)=(\pi,\pi)$}\label{sec:pipialpha}
  In this mean-field ansatz, we have obtained the total EM electric charge $N_E$ in the \emph{Standard Labeling} and in the presence of $\alpha$ (cf. Eq.(\ref{chargestandard1})). To approach a mirror-symmetric charge lattice, we  require that the site distribution and quantum statistics are mirror-symmetric.

\fbox{$l/s\in\mathbb{Z}$}. Let us first consider $l/s\in\mathbb{Z}$ such that $N_M$ is arbitrarily integer-valued due to the constraint Eq.(\ref{constraint1}). We assume that the mirror site of $(N_M,n_f^{(2)}, N^{(1)}_m)$ is labeled by $(-N_M, n_f^{(2)'}, N^{(1)'}_m)$. In order that the mirror site does exist in the charge latice, the integer solutions $(n_f^{(2)'}, N^{(1)'}_m)$ of the following equation must exist for any given integer $N_M$:
 \begin{align}
N_E =&-(\frac{l}{s}+1)N^{(1)'}_m+n^{(2)'}_f-(2\alpha \frac{l}{s}+\alpha+\frac{1}{2})N_M
\end{align}
which is equivalent to
 \begin{align}
&\left(n^{(2)'}_f-n^{(2)}_f\right)- (\frac{l}{s}+1)\left(N^{(1)'}_m- N^{(1)}_m\right)\nonumber\\
=&2(2\alpha \frac{l}{s} +\alpha+\frac{1}{2})N_M\,\label{nm3}
\end{align}
by means of Eq.(\ref{chargestandard1}).    Therefore, the mirror-symmetric site distribution requires that:  $2\times(2\alpha\frac{l}{s}+\alpha+\frac{1}{2})=\text{integer}$. To construct symmetric topological phases, we need further check the quantum statistics in the presence of $\alpha$ (details of derivation are present in Appendix. \ref{appendix_q2}):
\begin{align}
\Gamma=&N_M \left(N_E-(2\alpha\frac{l}{s}+\alpha+\frac{1}{2})N_M-(\frac{l}{s}+1)\right)\,.\label{q2}
\end{align}
Therefore, the quantum statistics is mirror-symmetric if
\begin{align}
&N_M \left(N_E-(2\alpha\frac{l}{s}+\alpha+\frac{1}{2})N_M-(\frac{l}{s}+1)\right)\,\nonumber\\
\doteq&-N_M \left(N_E+(2\alpha\frac{l}{s}+\alpha+\frac{1}{2})N_M-(\frac{l}{s}+1)\right)\,\nonumber\\
&\text{\emph{i.e.} }2N_M N_E\doteq 0\,.\label{alphacondition1}
\end{align}
In other words, $2N_M N_E$ must be always even.

If $2\times( 2\alpha\frac{l}{s}+\alpha+\frac{1}{2})=\text{odd}$, Eq.(\ref{chargestandard1}) indicates that $N_E$ is half-odd integer if we take $N_M=1$. As a result, Eq.(\ref{alphacondition1}) is not satisfied.  Therefore, in order to guarantee mirror-symmetric distribution of quantum statistics,  we need to consider a stronger condition:  $2\times(2\alpha\frac{l}{s}+\alpha+\frac{1}{2})=\text{even}$, \emph{i.e.} $(2\alpha\frac{l}{s}+\alpha+\frac{1}{2})=\text{integer}$.  In Eq.(\ref{eq:tot_chargepipi}), we have already proved that energy is mirror-symmetric for any $\alpha$, so that we conclude that the charge lattice is mirror-symmetric if the stronger condition ``$2\alpha\frac{l}{s}+\alpha+\frac{1}{2}=\text{integer}$'' is satisfied.

Since $N_M\in\mathbb{Z}$, $\frac{l}{s}\in\mathbb{Z}$ and $(\frac{l}{s}+1)N_M\doteq(\frac{l}{s}+1)(N_M)^2$, one may rewrite Eq.(\ref{q2}) as $\Gamma\doteq N_M(N_E-(2\alpha\frac{l}{s}+\alpha+\frac{1}{2})N_M-(\frac{l}{s}+1)N_M)\equiv N_M(N_E-\frac{\Theta}{2\pi}N_M)$. The minimal periodicity of $\Theta$ is $4\pi$ because $\Gamma$ is invariant after $4\pi$ shift. As a result, a $\Theta$ angle can be formally defined:
\begin{align}
{\Theta}\equiv 2\pi(\frac{l}{s}+\frac{3}{2})+4\pi(\frac{l}{s}+\frac{1}{2})\alpha\,\,\,\text{mod} (4\pi)\,\label{Thetapipi}
\end{align}
from which we see that the $\Theta$ angle is linearly related to $\alpha$.  

From this $\Theta$ formula, we realize that shifting $\alpha$ by odd integer will change trivial (non-trivial) SPT to non-trivial (trivial) SPT. Therefore, we arrive at the statement in Sec.\ref{axionfield}.   
Thus, two allowed values of $\alpha$ must be differed from each other by an even integer. Let us consider $\overline{\alpha}=1-\alpha$ and $\alpha$ where $\overline{\alpha}$ is the time-reversal transformed $\alpha$ shown in Eq.(\ref{alphabar}). The requirement $\overline{\alpha}-\alpha=\text{any even integer}$ is equivalent to the constraint $\alpha=\text{half-odd}$ which is nothing but Condition-\uppercase\expandafter{\romannumeral3}.  Under this condition as well as the conditions obtained from mirror symmetric charge lattice,   we may obtain   trivial SPT states and non-trivial SPT states summarized in Table \ref{tab:results1} by comparing $\Gamma$ with the standard trivial Witten effect and non-trivial Witten effect defined in Sec.\ref{sec:definingwitten}.  Strikingly, we obtain BTI states which are completely absent in Table \ref{tab:results} where $\alpha=1/2$ is fixed.

\fbox{$l/s\notin\mathbb{Z}$}. Generally, we may parametrize $l/s=k'+\frac{p}{q}$, where, $k',p,q\in\mathbb{Z}$, $q>p>0$, $\text{gcd}(p,q)=1$ (gcd: greatest common divisor). In this case, $N_M$ is quantized at $qk$ as shown in Eq.(\ref{quantizationnm1}).  To guarantee mirror-symmetric site distribution, the integer solutions $(n_f^{(2)'}, N^{(1)'}_m)$ of Eq.(\ref{nm3}) must exist for any given  $N_M=qk$:
 \begin{align}
&\left(n^{(2)'}_f-n^{(2)}_f\right)- (k'+1+\frac{p}{q})\left(N^{(1)'}_m- N^{(1)}_m\right)\nonumber\\
=&(4\alpha(qk'+p)+2\alpha q+q)k\,\label{nm4}
\end{align}
by means of Eq.(\ref{chargestandard1}).  Eq.(\ref{q2}) is also valid when $l/s\notin\mathbb{Z}$ by noting that $-2N_M(\frac{l}{s}+1)(N^{(1)}_m+1)$ is still even integer in deriving the fourth line of Appendix \ref{appendix_q2}.  Therefore, Eq.(\ref{alphacondition1}) is also valid when $l/s\notin\mathbb{Z}$.

A general discussion on Eq.(\ref{nm4}) and Eq.(\ref{alphacondition1}) is intricate. Let us take a simple example: $l/s=1/3$, \emph{i.e.} $k'=0,q=3,p=1$. The right hand side of Eq.(\ref{nm4}) becomes $(10\alpha+3) k$. To obtain the integer solutions $(n_f^{(2)'}, N^{(1)'}_m)$ for any given  integers ($k, N_m^{(1)}, n^{(2)}_f$),  a constraint on $\alpha$ is necessary: $10\alpha+3={\text{integer}}/{3}$, \emph{i.e.} $\alpha=\frac{k_0-9}{30}$ where $k_0$ is an integer. Under this condition, Eq.(\ref{alphacondition1}) leads to a stronger condition: $\alpha=\frac{2k_0-9}{30}$ which guarantees mirror-symmetric distributions of both sites and quantum statistics.  As we have proved, excitation energy is already mirror-symmetric due to Eq.(\ref{eq:tot_chargepipi}). By further considering the Condition-\uppercase\expandafter{\romannumeral3}, $\alpha$ is finally restricted to: $\alpha=\text{half-odd}$.

Since $N_M/3\in\mathbb{Z}$, $\frac{l}{s}=1/3$ and $(\frac{l}{s}+1)N_M=4N_M/3\doteq 0$, one may rewrite Eq.(\ref{q2}) as $\Gamma\doteq N_M(N_E-(\frac{5\alpha}{3}+\frac{1}{2})N_M)\equiv N_M(N_E-\frac{\Theta}{2\pi}N_M)$. The minimal periodicity of $\Theta$ is $\frac{4\pi}{9}$ because $\Gamma$ is invariant after $\frac{4\pi}{9}$ shift. As a result, a $\Theta$ angle can be formally defined: 
\begin{align}
{\Theta}\equiv \pi +\frac{10\pi}{3}\alpha\,\,\,\text{mod} (\frac{4\pi}{9})\,\label{Thetapipifractional}
\end{align}
from which we see that the $\Theta$ angle is linearly related to $\alpha$.   All SET states have fractional intrinsic excitations. We  can further classify these states into two categories: one is $\Theta=0\,\,\,\text{mod} (\frac{4\pi}{9})$ with $\alpha-\frac{1}{2}=\text{even}$; one is $\Theta=\frac{2\pi}{9}\,\,\,\text{mod} (\frac{4\pi}{9})$ with $\alpha-\frac{1}{2}=\text{odd}$. In comparison to the trivial and non-trivial SPT states, we call the former ``trivial \emph{f}BTI'' and the latter ``non-trivial \emph{f}BTI'' via investigating the Witten effect.

   As usual, one should pay attention to the emergence of Z$_{|s|}$ TO if $l/s=-1/2$.  One may also examine whether there is a Witten effect if $l/s=-1/2$ in addition to Z$_{|s|}$ TO. Following the same procedure, we obtain that:
\begin{align}
{\Theta}\equiv \frac{\pi}{2}  \,\,\,\text{mod} (\pi)\,\label{Thetapipispecial}
\end{align}
and $\alpha$ is restricted to $\alpha=\text{half-odd}$. At this special point $l/s=-1/2$, we find that Witten effect is independent on $\alpha$ and the state is a non-trivial \emph{f}BTI in the presence of Z$_{|s|}$ TO.

\section{Conclusion}\label{sec:conclusion}

In conclusion, we use fermionic projective construction and dyon condensation
to construct many three-dimensional SPT and SET states with time-reversal
symmetry and U(1) boson number conservation symmetry.  Without dyon
condensation, we have an algebraic bosonic insulator which contains an emergent
U(1) gapless photon excitation. Then we assume the internal U(1) gauge field to
fluctuate strongly and form one of many confined phases characterized by
different dyon condensations.  After a dyon condensate that preserve the
$U(1)\rtimes Z^T_2$ symmetry is selected properly, the excitation spectrum
(formed by deconfined dyons) above this dyon condensate is entirely fixed. The
symmetric dyon condensate determines the quantization conditions of EM magnetic
charge and EM electric charge of excitations.  It also determines the quantum
statistics (boson/fermion) and excitation energy.  By calculating these
properties, we then obtain SPT and SET states summarized in Tables
\ref{tab:results0}, \ref{tab:results}, and \ref{tab:results1}. The basic
process of this construction approach is shown in Fig.  \ref{figure_process}.  

There are some interesting and direct directions for future work.

\emph{(1) Classification via projective construction and dyon condensation.}
The definition of non-trivial SPT states, \emph{i.e.} bosonic topological
insulators (BTI), is only related to the non-trivial Witten effect (\emph{i.e.}
$\Theta=2\pi$) as shown in Sec.\ref{sec:definingwitten}. As shown in
Ref.\onlinecite{Wen13topo}, classification of a SPT state with a certain
symmetry corresponds to looking for a complete set of ``topological
invariants''. In this paper, we only consider one Z$_2$ topological invariant
which distinguishes the physical properties of trivial/non-trivial Witten
effect, meaning that it is potentially possible  some trivial SPT states  we found in this
paper actually are non-trivial via other physical properties that can not be
realized in a truly trivial Mott insulator state. In other words, it is necessary to 
construct more topological invariants to \emph{completely} distinguish all  SPT
states. 

 There are some clues. Firstly, in this  paper, we have systematically
shown how to construct a charge lattice which respects symmetry by means of  fermionic projective
construction and dyon condensation.  We expect that, in addition to Witten
effect,  more information (\emph{i.e.} more topological invariants)  can be
extracted from more complete analysis of charge lattice. Secondly, we may
consider   SPT states with merely time-reversal symmetry. In other words,
these states are protected sufficiently by time-reversal symmetry while the
boson number conservation symmetry U(1) doesn't play any role. Literally, these
states are also SPT states with U(1)$\rtimes$Z$^T_2$. Therefore, one may
consider new mean-field ansatzes for fermions and try to find new SPT states. 

\emph{(2) Surface theory and bulk topological field theory via projective construction and dyon condensation.}  SPT states have quite trivial bulk but the surface may admit many non-trivial physical properties than is absent in trivial Mott insulator states. It has been recently shown that classifying surface topological order may provide the answer to classifying the BTI bulk.\cite{VS1258,WT13}  Indeed, the surface detectable features may be tightly connected to the complete set of topological invariants that we shall look for. For instance, a non-trivial Witten effect is indeed related to the surface quantum Hall effect (by breaking time-reversal symmetry on the surface) with  anomalous quantization of Hall conductance that cannot be realized in 2D U(1) SPT.\cite{WT13,Ye13long} In short, it is interesting for future work on surface theoretical description via the present fermionic projective construction and dyon construction.  Beside the SPT states constructed in this paper, we also constructed many SET states in which fractional intrinsic excitations (defined as excitations with zero EM magnetic charge). A full understanding on these topologically ordered states with symmetry is interesting, such as (3+1)D topological quantum field theory (TQFT) descriptions, fixed-point lattice Hamiltonian realization.

%
%
%

\section*{Acknowledgement} 
We would like to thank Max Metlitski, G. Baskaran and Sung-Sik Lee for many helpful
discussions.  This research is supported by NSF Grant No.  DMR-1005541, NSFC
11074140, and NSFC 11274192. (X.G.W.) Research at Perimeter Institute is
supported by the Government of Canada through Industry Canada and by the
Province of Ontario through the Ministry of Research and Innovation.  (P.Y. and
X.G.W.)

\appendix
 
\section{Condensation of two time-reversal conjugated dyons leads to single dyon condensate}\label{appendix_twodyons}

\begin{widetext}
The most general ansatz for the quantum numbers of two time-reversal conjugated dyons is shown in Table. \ref{tab:twodyons}. We see that there are four numbers $(l_1,l_2,s,t)$ enough to labels two dyons one of which is the time-reversal partner of the other.

\begin{table}
\centering
\caption{Quantum numbers of two time-reversal conjugated dyons ($\phi_1$ and $\phi_2$)}
\label{tab:twodyons}
\begin{tabular}{|c|c|c|c|c|c|c|c|}\hline
Dyon&$N_m^a$&$N_f^{(1)}$&$N_f^{(2)}$&$N_m^{(1)}$&$N_m^{(2)}$&$N_M$&$N_A$\\ \hline
$\phi_1$&$s$&$l_1$&$l_2$&$s+\alpha t$&$-s+(1-\alpha)t$&$t$&$\alpha l_1+(1-\alpha)l_2$\\ \hline
$\phi_2$&$s+(2\alpha-1)t$&$l_2$&$l_1$&$s-(1-\alpha)t$&$-s-\alpha t$&$-t$&$\alpha l_2+(1-\alpha)l_1$\\ 
\hline
\hline
\end{tabular}
\end{table}

If both $\phi_1$ and $\phi_2$ are condensed, the mutual statistics between them must be trivialized, \emph{i.e.}
\begin{align}
&l_1 (s-(1-\alpha)t)+l_2 (-s-\alpha t)\nonumber\\
=&l_2 (s+\alpha t) +l_1 (-s+(1-\alpha)t)\,
\end{align}
which leads to
\begin{align}
l_1 (s-(1-\alpha)t)+l_2 (-s-\alpha t)=0\,.
\end{align}
We note that the U(1) symmetry of the original boson system is generated by conserved EM electric charge rather than EM magnetic charge. The EM field here is non-dynamical so that the dyon condensates which form the physical ground state should have zero EM magnetic charge. Therefore, we have $t=0$, and the above trivial mutual statistics condition becomes:
 \begin{align}
(l_1 -l_2) s=0\,.
\end{align}
To satisfy this condition, $l_1=l_2$ or $s=0$. But $s$ must be nonzero in order to preserve U(1) symmetry. The reason is that once $s=0$, the condensation will break U(1) since it is EM electric charged. The reasonable choice is $l_1=l_2$. This choice leads to the fact that $\phi_1=\phi_2$, \emph{i.e.} a condensate of one kind of dyon. 
\end{widetext}

\section{Derivation of Eq.(\ref{chargestandard})}\label{appendix_eq:totalcharge}
\begin{align}
N_E=&(\alpha N^{(1)}_f+(1-\alpha)N^{(2)}_f)-\frac{l}{s}N_m^a\nonumber\\   
 =& \alpha n^{(1)}_f+(1-\alpha)n^{(2)}_f -\frac{l}{s}(N^{(1)}_m-\alpha N_M)\nonumber\\  
 =&\alpha (\frac{l}{s}N_M+n^{(2)}_f)+(1-\alpha)n^{(2)}_f+\frac{l}{s}\alpha N_M- \frac{l}{s}N^{(1)}_m\,\nonumber\\
 =& n^{(2)}_f+ 2\alpha \frac{l}{s} N_M- \frac{l}{s}N^{(1)}_m\,.   
 \end{align}
In deriving the first line, Eqs.(\ref{na},\ref{eq:totalcharge}) are applied. In deriving the second line,   Eq.(\ref{eq_quantization})  is applied. In deriving the third line, Eq. (\ref{constraint}) is applied.

\section{Derivation of Eq.(\ref{eq:excitation_sign})}\label{appendix_eq:excitation_sign}
\begin{align}
\Gamma\doteq&N_M (\frac{1}{2}N^{(1)}_f+\frac{1}{2}N^{(2)}_f)+N^a_m (N^{(1)}_f-N^{(2)}_f)+N^{(1)}_f+N^{(2)}_f\nonumber\\
\doteq&N_M (\frac{1}{2}N^{(1)}_f+\frac{1}{2}N^{(2)}_f)+N^a_m (N^{(1)}_f-N^{(2)}_f)+N^{(1)}_f-N^{(2)}_f\nonumber\\
=&N_M (\frac{1}{2}N^{(1)}_f+\frac{1}{2}N^{(2)}_f)+(N^a_m+1) (N^{(1)}_f-N^{(2)}_f)\nonumber\\
=&N_MN_A+(N^a_m+1) N^a\nonumber\\
=&N_M(N_E+\frac{l}{s}N^a_m)+(N^a_m+1) \frac{l}{s}N_M\nonumber\\
=&N_MN_E+(2N^a_m+1) \frac{l}{s}N_M\nonumber\\
=&N_MN_E+(2N^{(1)}_m-N_M+1) \frac{l}{s}N_M
\end{align}
In deriving the second line, an even integer ``$-2N^{(2)}_f$'' is added. In deriving the fifth line, Eqs.(\ref{constraint}, \ref{eq:totalcharge}) are applied. In deriving the last line, Eq.(\ref{eq_quantization}) is applied.

\section{Derivation of Eq.(\ref{chargestandard1})}\label{appendix_chargestandard1}

\begin{align}
N_E=&N_A-\frac{l+\frac{s}{2}}{s}N_m^a\nonumber\\
=&\alpha\left(n^{(1)}_f+\frac{1}{2}N_m^{(1)}\right)  +(1-\alpha)\left(n^{(2)}_f+\frac{1}{2}(N_M-N_m^{(1)})\right)\nonumber\\
&-(\frac{l}{s}+\frac{1}{2})(  N^{(1)}_m-\alpha N_M)\nonumber\\
=&\alpha\left(    n^{(2)}_f+(\frac{l}{s}+1)N_M-N_m^{(1)}+\frac{1}{2}N_m^{(1)}\right) \nonumber\\
& +(1-\alpha)\left(n^{(2)}_f+\frac{1}{2}(N_M-N_m^{(1)})\right)\nonumber\\
&-(\frac{l}{s}+\frac{1}{2})(  N^{(1)}_m- \alpha N_M)\nonumber\\
=&-(\frac{l}{s}+1)N^{(1)}_m+n^{(2)}_f+(2\alpha \frac{l}{s}+\alpha+\frac{1}{2})N_M
\end{align}
In deriving the third line, Eq.(\ref{constraint1}) is applied.

\section{Derivation of Eq.(\ref{eq:excitation_sign1})} \label{appendix_eq:excitation_sign1}

\begin{align}
\Gamma\doteq~&\frac{1}{2}N_M (n^{(1)}_f+n^{(2)}_f)+N^a_m (n^{(1)}_f-n^{(2)}_f)+n^{(1)}_f+n^{(2)}_f\nonumber\\
\doteq&\frac{1}{2}N_M (n^{(1)}_f+n^{(2)}_f)+N^a_m (n^{(1)}_f-n^{(2)}_f)+n^{(1)}_f-n^{(2)}_f\nonumber\\
=&(\frac{1}{2}N_M +N_m^a+1)(n^{(1)}_f-n^{(2)}_f)+N_M n^{(2)}_f\nonumber\\
=&(N^{(1)}_m+1)(n^{(1)}_f-n^{(2)}_f)+N_M n^{(2)}_f\nonumber\\
=&(N^{(1)}_m+1)\left((\frac{l}{s}+1)N_M-N^{(1)}_m\right)+N_M n^{(2)}_f\nonumber\\
\doteq&(N^{(1)}_m+1)(\frac{l}{s}+1)N_M+N_M n^{(2)}_f
\end{align}
In deriving the first line, Eq.(\ref{eqn:stat1}) is applied. In deriving the second line, an even integer ``$-2n^{(2)}_f$'' is added. In deriving the fourth line, the first formula in Eq.(\ref{eq_quantization}) is applied with $\alpha=1/2$. In deriving the fifth line, Eq.(\ref{constraint1}) is applied. In deriving the last line, the even integer ``$-N^{(1)}_m (N^{(1)}_m+1)$'' is removed.

\section{Derivation of Eq.(\ref{eqa})}\label{appendix_eqa}
\begin{align}
\Gamma\doteq& N_M(N^{(1)}_m+1)(\frac{l}{s}+1)+N_M n^{(2)}_f\nonumber\\
 \doteq& -N_M(N^{(1)}_m+1)(\frac{l}{s}+1)+N_M n^{(2)}_f\nonumber\\
=&N_M\left(-(\frac{l}{s}+1)N^{(1)}_m+ n^{(2)}_f-(\frac{l}{s}+1)\right)\nonumber\\
=&N_M\left(N_E-(\frac{l}{s}+1)N_M-(\frac{l}{s}+1)\right)\nonumber\\
=&N_M\left(N_E-(\frac{l}{s}+1)(N_M+1)\right)\nonumber\\
\doteq&N_M N_E 
 \end{align}
where, an even integer ``$-2N_M (N^{(1)}_m+1)(\frac{l}{s}+1)$'' is added in the second line. In the fourth line, Eq.(\ref{eq:tot_charge2}) is applied. In the last line, ``$N_M(N_M+1)(\frac{l}{s}+1)$'' is removed since it is always even.

\section{Derivaion of Eq.(\ref{eqa1})}\label{appendix_eqa1}
  \begin{align}
\Gamma\doteq& N_M(N^{(1)}_m+1)(\frac{l}{s}+1)+N_M n^{(2)}_f\nonumber\\
 \doteq& -N_M(N^{(1)}_m+1)(\frac{l}{s}+1)+N_M n^{(2)}_f\nonumber\\
=&N_M\left(-(\frac{l}{s}+1)N^{(1)}_m+ n^{(2)}_f-(\frac{l}{s}+1)\right)\nonumber\\
=&N_M\left(N_E-(\frac{l}{s}+1)N_M-(\frac{l}{s}+1)\right)\nonumber\\
=&N_MN_E-(\frac{l}{s}+1)N_M(N_M+1) \nonumber\\
\doteq&N_MN_E- \frac{l}{s}N_M(N_M+1)\nonumber\\
\doteq&N_MN_E-\frac{l}{s} N_M(N_M-1) 
 \end{align}
 where, an even integer ``$-2N_M (N^{(1)}_m+1)(\frac{l}{s}+1)=-2qk(N^{(1)}_m+1)(\frac{p}{q}+1)=-2k(N^{(1)}_m+1)(p+q)$'' is added in the second line. In deriving the  fourth line, Eq.(\ref{eq:tot_charge2}) is applied. In deriving the sixth line, the even integer ``$-N_M(N_M+1)=-qk (qk+1)$'' is removed. In deriving the last line, an even integer ``$2\frac{l}{s}N_M=2(qk'+p)k$'' is added.

 \section{Derivation of Eq.(\ref{q1})}\label{appendix_q1}
\begin{align}
\Gamma=&(N^{(1)}_m+1)n^{(1)}_f+(N^{(2)}_m+1)n^{(2)}_f\nonumber\\
=&(N^{(1)}_m+1)(\frac{l}{s}N_M+n^{(2)}_f)+(N_M-N^{(1)}_m+1)n^{(2)}_f\nonumber\\
\doteq&N_M(\frac{l}{s}N^{(1)}_m+\frac{l}{s}+n^{(2)}_f)\nonumber\\
\doteq&N_M(-\frac{l}{s}N^{(1)}_m+\frac{l}{s}+n^{(2)}_f)\nonumber\\
=&N_M \left(N_E-2\alpha\frac{l}{s}N_M+\frac{l}{s}\right)\,.
\end{align}
In deriving the second line, Eq.(\ref{constraint}) and Eq.(\ref{combine_NM}) are applied. In deriving the third line, the even integer $2n^{(2)}_f$ is removed. In deriving the fourth line, an even integer $-2N_M\frac{l}{s}N^{(1)}_m$ is added. In deriving the last line, Eq.(\ref{chargestandard}) is applied.

\section{Derivation of Eq.(\ref{q2})}\label{appendix_q2}
\begin{align}
\Gamma=&(N^{(1)}_m+1)n^{(1)}_f+(N^{(2)}_m+1)n^{(2)}_f\nonumber\\
=&(N^{(1)}_m+1)\left((\frac{l}{s}+1)N_M-N^{(1)}_m+n^{(2)}_f\right)\nonumber\\
&+(N_M-N^{(1)}_m+1)n^{(2)}_f\nonumber\\
\doteq&N_M\left((\frac{l}{s}+1)(N^{(1)}_m+1)+n^{(2)}_f\right)\nonumber\\
\doteq&N_M\left(-(\frac{l}{s}+1)(N^{(1)}_m+1)+n^{(2)}_f\right)\nonumber\\
=&N_M \left(N_E-(2\alpha\frac{l}{s}+\alpha+\frac{1}{2})N_M-(\frac{l}{s}+1)\right)\,.
\end{align}
In deriving the second line, Eq.(\ref{constraint1}) and Eq.(\ref{combine_NM}) are applied. In deriving the third line, the even integers $2n^{(2)}_f$ and  $-(N^{(1)}_m+1)N^{(1)}_m$ are removed. In deriving the fourth line, an even integer $-2N_M(\frac{l}{s}+1)(N^{(1)}_m+1)$ is added. In deriving the last line, Eq.(\ref{chargestandard1}) is applied.


\begin{thebibliography}{99}
\expandafter\ifx\csname natexlab\endcsname\relax\def\natexlab#1{#1}\fi
\expandafter\ifx\csname bibnamefont\endcsname\relax
  \def\bibnamefont#1{#1}\fi
\expandafter\ifx\csname bibfnamefont\endcsname\relax
  \def\bibfnamefont#1{#1}\fi
\expandafter\ifx\csname citenamefont\endcsname\relax
  \def\citenamefont#1{#1}\fi
\expandafter\ifx\csname url\endcsname\relax
  \def\url#1{\texttt{#1}}\fi
\expandafter\ifx\csname urlprefix\endcsname\relax\def\urlprefix{URL }\fi
\providecommand{\bibinfo}[2]{#2}
\providecommand{\eprint}[2][]{\url{#2}}



\bibitem[{\citenamefont{Wen}(1989)}]{Wtop}
\bibinfo{author}{\bibfnamefont{X.-G.} \bibnamefont{Wen}},
  \bibinfo{journal}{Phys. Rev. B} \textbf{\bibinfo{volume}{40}},
  \bibinfo{pages}{7387} (\bibinfo{year}{1989}).

\bibitem[{\citenamefont{Wen and Niu}(1990)}]{WNtop}
\bibinfo{author}{\bibfnamefont{X.-G.} \bibnamefont{Wen}} \bibnamefont{and}
  \bibinfo{author}{\bibfnamefont{Q.}~\bibnamefont{Niu}},
  \bibinfo{journal}{Phys. Rev. B} \textbf{\bibinfo{volume}{41}},
  \bibinfo{pages}{9377} (\bibinfo{year}{1990}).

\bibitem[{\citenamefont{Wen}(1990)}]{Wrig}
\bibinfo{author}{\bibfnamefont{X.-G.} \bibnamefont{Wen}},
  \bibinfo{journal}{Int. J. Mod. Phys. B} \textbf{\bibinfo{volume}{4}},
  \bibinfo{pages}{239} (\bibinfo{year}{1990}).

\bibitem{H8483}%
B. I. Halperin,
Phys. Rev. Lett., {\bf 52}, 1583 (1984).

\bibitem{ASW8422}%
  \bibinfo {author} {\bibfnamefont{D.}~\bibnamefont{Arovas}}, \bibinfo {author}
  {\bibfnamefont{J.~R.}\ \bibnamefont{Schrieffer}},\ and\ \bibinfo {author}
  {\bibfnamefont{F.}~\bibnamefont{Wilczek}},
  \bibinfo {journal} {Phys. Rev. Lett.}
  \textbf{\bibinfo {volume} {53}},\ \bibinfo {pages} {722} (\bibinfo {year}
  {1984})%

\bibitem{H8285}%
  \bibinfo {author} {\bibfnamefont{B.~I.}\ \bibnamefont{Halperin}},
  \bibinfo {journal} {Phys. Rev. B}
  \textbf{\bibinfo {volume} {25}},\ \bibinfo {pages} {2185} (\bibinfo {year}
  {1982})%

\bibitem{Wedgerev}%
  \bibinfo {author} {\bibfnamefont{X.-G.}\ \bibnamefont{Wen}},
  \bibinfo {journal} {Int. J. Mod. Phys. B}
  \textbf{\bibinfo {volume} {6}},\ \bibinfo {pages} {1711} (\bibinfo {year}
  {1992})%

\bibitem{FNS0428}%
  \bibinfo {author} {\bibfnamefont{M.}~\bibnamefont{Freedman}}, \bibinfo
  {author} {\bibfnamefont{C.}~\bibnamefont{Nayak}}, \bibinfo {author}
  {\bibfnamefont{K.}~\bibnamefont{Shtengel}}, \bibinfo {author}
  {\bibfnamefont{K.}~\bibnamefont{Walker}},\ and\ \bibinfo {author}
  {\bibfnamefont{Z.}~\bibnamefont{Wang}},
  \bibinfo {journal} {Ann. Phys. (NY)}
  \textbf{\bibinfo {volume} {310}},\ \bibinfo {pages} {428} (\bibinfo {year}
  {2004})%

\bibitem{LW0510}%
  \bibinfo {author} {\bibfnamefont{M.}~\bibnamefont{Levin}}\ and\ \bibinfo
  {author} {\bibfnamefont{X.-G.}\ \bibnamefont{Wen}},
  \bibinfo {journal} {Phys. Rev. B}
  \textbf{\bibinfo {volume} {71}},\ \bibinfo {pages} {045110} (\bibinfo {year}
  {2005}),\ \eprint{http://arxiv.org/abs/cond-mat/0404617}{cond-mat/0404617}%


\bibitem{Chenlong} Xie Chen, Z. C. Gu, Z. X. Liu, and X. G. Wen, arxiv:1106.4772.

\bibitem{Chen10} Xie Chen, Z. C. Gu, and X. G. Wen, Phys. Rev. B \textbf{82}, 155138 (2010).

  \bibitem{Wenscience} X. Chen, Z.-C. Gu, Z.-X. Liu, X.-G. Wen, Science 338, 1604 (2012); X. L. Qi, Science 338, 1550 (2012).

\bibitem[{\citenamefont{Levin and Stern}(2009)}]{LS0903}
\bibinfo{author}{\bibfnamefont{M.}~\bibnamefont{Levin}} \bibnamefont{and}
  \bibinfo{author}{\bibfnamefont{A.}~\bibnamefont{Stern}},
  \bibinfo{journal}{Phys. Rev. Lett.} \textbf{\bibinfo{volume}{103}},
  \bibinfo{pages}{196803} (\bibinfo{year}{2009}), \eprint{arXiv:0906.2769}.

\bibitem[{\citenamefont{Lu and Vishwanath}(2012)}]{LV1219}
\bibinfo{author}{\bibfnamefont{Y.-M.} \bibnamefont{Lu}} \bibnamefont{and}
  \bibinfo{author}{\bibfnamefont{A.}~\bibnamefont{Vishwanath}},
  \bibinfo{journal}{Phys. Rev.} \textbf{\bibinfo{volume}{86}},
  \bibinfo{pages}{125119} (\bibinfo{year}{2012}), \eprint{arXiv:1205.3156}.

\bibitem{VS1258} Ashvin Vishwanath, T. Senthil, Phys. Rev. X 3, 011016 (2013), arXiv:1209.3058; 
 
 
 \bibitem{Burnell13} F. J. Burnell, X. Chen, L. Fidkowski, and A. Vishwanath (2013), arXiv:1302.7072.

\bibitem{Cheng13} Meng Cheng and Zheng-Cheng Gu (2013), arXiv: 1302.4802.

\bibitem{Chen13} Xie Chen, Fa Wang, Y.-M. Lu, and D.-H. Lee (2013), arXiv:1302.3121.

\bibitem{Wen13topo} X.-G. Wen (2013), arXiv: 1301.7675.
\bibitem{Xu13} Cenke Xu (2013), arXiv: 1301.6172.

\bibitem{MM13} M. A. Metlitski, C. L. Kane, and M. P. A. Fisher, arXiv: 1302.6535 (2013).

\bibitem{MM13a} M. A. Metlitski, C. L. Kane, and M. P. A. Fisher, to appear.

\bibitem{Hung1211} Ling-Yan Hung and X.-G. Wen (2013), arXiv:1211.2767.
\bibitem{Senthillevin12} T. Senthil and M. Levin (2013). arXiv:1206.1604.
\bibitem{Xu12} Cenke Xu (2012), arXiv:1209.4399.  
\bibitem[{\citenamefont{Baskaran et~al.}(1987)\citenamefont{Baskaran, Zou, and
  Anderson}}]{BZA8773}
\bibinfo{author}{\bibfnamefont{G.}~\bibnamefont{Baskaran}},
  \bibinfo{author}{\bibfnamefont{Z.}~\bibnamefont{Zou}}, \bibnamefont{and}
  \bibinfo{author}{\bibfnamefont{P.~W.} \bibnamefont{Anderson}},
  \bibinfo{journal}{Solid State Comm.} \textbf{\bibinfo{volume}{63}},
  \bibinfo{pages}{973} (\bibinfo{year}{1987}).

\bibitem[{\citenamefont{Baskaran and Anderson}(1988)}]{BA8880}
\bibinfo{author}{\bibfnamefont{G.}~\bibnamefont{Baskaran}} \bibnamefont{and}
  \bibinfo{author}{\bibfnamefont{P.~W.} \bibnamefont{Anderson}},
  \bibinfo{journal}{Phys. Rev. B} \textbf{\bibinfo{volume}{37}},
  \bibinfo{pages}{580} (\bibinfo{year}{1988}).

\bibitem[{\citenamefont{Affleck and Marston}(1988)}]{AM8874}
\bibinfo{author}{\bibfnamefont{I.}~\bibnamefont{Affleck}} \bibnamefont{and}
  \bibinfo{author}{\bibfnamefont{J.~B.} \bibnamefont{Marston}},
  \bibinfo{journal}{Phys. Rev. B} \textbf{\bibinfo{volume}{37}},
  \bibinfo{pages}{3774} (\bibinfo{year}{1988}).

\bibitem[{\citenamefont{Kotliar and Liu}(1988)}]{KL8842}
\bibinfo{author}{\bibfnamefont{G.}~\bibnamefont{Kotliar}} \bibnamefont{and}
  \bibinfo{author}{\bibfnamefont{J.}~\bibnamefont{Liu}},
  \bibinfo{journal}{Phys. Rev. B} \textbf{\bibinfo{volume}{38}},
  \bibinfo{pages}{5142} (\bibinfo{year}{1988}).

\bibitem[{\citenamefont{Suzumura et~al.}(1988)\citenamefont{Suzumura, Hasegawa,
  and Fukuyama}}]{SHF8868}
\bibinfo{author}{\bibfnamefont{Y.}~\bibnamefont{Suzumura}},
  \bibinfo{author}{\bibfnamefont{Y.}~\bibnamefont{Hasegawa}}, \bibnamefont{and}
  \bibinfo{author}{\bibfnamefont{H.}~\bibnamefont{Fukuyama}},
  \bibinfo{journal}{J. Phys. Soc. Jpn.} \textbf{\bibinfo{volume}{57}},
  \bibinfo{pages}{2768} (\bibinfo{year}{1988}).

\bibitem[{\citenamefont{Affleck et~al.}(1988)\citenamefont{Affleck, Zou, Hsu,
  and Anderson}}]{AZH8845}
\bibinfo{author}{\bibfnamefont{I.}~\bibnamefont{Affleck}},
  \bibinfo{author}{\bibfnamefont{Z.}~\bibnamefont{Zou}},
  \bibinfo{author}{\bibfnamefont{T.}~\bibnamefont{Hsu}}, \bibnamefont{and}
  \bibinfo{author}{\bibfnamefont{P.~W.} \bibnamefont{Anderson}},
  \bibinfo{journal}{Phys. Rev. B} \textbf{\bibinfo{volume}{38}},
  \bibinfo{pages}{745} (\bibinfo{year}{1988}).

\bibitem[{\citenamefont{Dagotto et~al.}(1988)\citenamefont{Dagotto, Fradkin,
  and Moreo}}]{DFM8826}
\bibinfo{author}{\bibfnamefont{E.}~\bibnamefont{Dagotto}},
  \bibinfo{author}{\bibfnamefont{E.}~\bibnamefont{Fradkin}}, \bibnamefont{and}
  \bibinfo{author}{\bibfnamefont{A.}~\bibnamefont{Moreo}},
  \bibinfo{journal}{Phys. Rev. B} \textbf{\bibinfo{volume}{38}},
  \bibinfo{pages}{2926} (\bibinfo{year}{1988}).

\bibitem[{\citenamefont{Wen et~al.}(1989)\citenamefont{Wen, Wilczek, and
  Zee}}]{WWZcsp}
\bibinfo{author}{\bibfnamefont{X.-G.} \bibnamefont{Wen}},
  \bibinfo{author}{\bibfnamefont{F.}~\bibnamefont{Wilczek}}, \bibnamefont{and}
  \bibinfo{author}{\bibfnamefont{A.}~\bibnamefont{Zee}},
  \bibinfo{journal}{Phys. Rev. B} \textbf{\bibinfo{volume}{39}},
  \bibinfo{pages}{11413} (\bibinfo{year}{1989}).

\bibitem[{\citenamefont{Wen}(1991)}]{Wsrvb}
\bibinfo{author}{\bibfnamefont{X.-G.} \bibnamefont{Wen}},
  \bibinfo{journal}{Phys. Rev. B} \textbf{\bibinfo{volume}{44}},
  \bibinfo{pages}{2664} (\bibinfo{year}{1991}).

\bibitem[{\citenamefont{Lee and Nagaosa}(1992)}]{LN9221}
\bibinfo{author}{\bibfnamefont{P.~A.} \bibnamefont{Lee}} \bibnamefont{and}
  \bibinfo{author}{\bibfnamefont{N.}~\bibnamefont{Nagaosa}},
  \bibinfo{journal}{Phys. Rev. B} \textbf{\bibinfo{volume}{45}},
  \bibinfo{pages}{5621} (\bibinfo{year}{1992}).

\bibitem[{\citenamefont{Mudry and Fradkin}(1994)}]{MF9400}
\bibinfo{author}{\bibfnamefont{C.}~\bibnamefont{Mudry}} \bibnamefont{and}
  \bibinfo{author}{\bibfnamefont{E.}~\bibnamefont{Fradkin}},
  \bibinfo{journal}{Phys. Rev. B} \textbf{\bibinfo{volume}{49}},
  \bibinfo{pages}{5200} (\bibinfo{year}{1994}).

\bibitem[{\citenamefont{Wen and Lee}(1996)}]{WLsu2}
\bibinfo{author}{\bibfnamefont{X.-G.} \bibnamefont{Wen}} \bibnamefont{and}
  \bibinfo{author}{\bibfnamefont{P.~A.} \bibnamefont{Lee}},
  \bibinfo{journal}{Phys. Rev. Lett.} \textbf{\bibinfo{volume}{76}},
  \bibinfo{pages}{503} (\bibinfo{year}{1996}), \eprint{cond-mat/9506065}.
  

%
%

\bibitem[{\citenamefont{Grover and Vishwanath}(2012)}]{GV1207}
\bibinfo{author}{\bibfnamefont{T.}~\bibnamefont{Grover}} \bibnamefont{and}
  \bibinfo{author}{\bibfnamefont{A.}~\bibnamefont{Vishwanath}}
  (\bibinfo{year}{2012}), \eprint{arXiv:1210.0907}.
\bibitem{Lu12b} Y.-M. Lu and D.-H. Lee (2012), arXiv:1210.0909.
\bibitem[{\citenamefont{Lu and Lee}(2012)}]{LL1263}
\bibinfo{author}{\bibfnamefont{Y.-M.} \bibnamefont{Lu}} \bibnamefont{and}
  \bibinfo{author}{\bibfnamefont{D.-H.} \bibnamefont{Lee}}
  (\bibinfo{year}{2012}), \eprint{arXiv:1212.0863}.
  
  
  \bibitem{YW12}Peng Ye and Xiao-Gang Wen, arXiv:1212.2121 (2012).


   \bibitem{TI}  J. E. Moore and L. Balents, Phys. Rev. B \textbf{75}, 121306 (2007); L. Fu, C. L. Kane, and E. J. Mele,
Phys. Rev. Lett. \textbf{98}, 106803 (2007); R. Roy, Phys. Rev. B, \textbf{79}, 195322 (2009); M. Z. Hasan, C. L. Kane, Rev. Mod. Phys. \textbf{82}, 3045
(2010); J. E. Moore, Nature \textbf{464}, 194 (2010); X.-L. Qi, S.-C. Zhang, Rev. Mod. Phys. \textbf{83}, 1057 (2011).

  \bibitem{Swingle10} B. Swingle,  M. Barkeshli, J. McGreevy, and T. Senthil, Phys. Rev. B \textbf{83}, 195139 (2011), arXiv:1005.1076 (2010).

  \bibitem{Maciejko10} J. Maciejko, X. L. Qi, A. Karch, S. C. Zhang, Phys. Rev. Lett. \textbf{105}, 246809 (2010).

\bibitem{BTI}Cenke Xu and T. Senthil, (2013). arXiv:1301.6172.

\bibitem{WS1334}C. Wang and T. Senthil, (2013). arXiv:1302.6234.

 \bibitem{Witten79} E. Witten, Phys. Lett. B \textbf{86}, 283 (1979).

 \bibitem{RF10} G. Rosenberg and M. Franz, Phys. Rev. B \textbf{82}, 035105 (2010).

 \bibitem{Goldhaber89} A. S. Goldhaber, R. MacKenzie, and F. Wilczek, Mod. Phys. Lett. A, \textbf{4}, 21 (1989).
\bibitem{Ye13long} Peng Ye and Juven Wang, arXiv:1306.3695.


%
%

\bibitem{WT13} C. Wang and T. Senthil, Phys. Rev. B \textbf{87}, 235122 (2013).
 \end{thebibliography}
\end{document}